\documentclass{article}
\usepackage[utf8]{inputenc}
\usepackage{parskip}
\usepackage{graphicx}
\usepackage{breqn}
\usepackage{amssymb}
\graphicspath{ {images/} }
\usepackage[a4paper, margin=1in]{geometry}
\usepackage[english]{babel}

\usepackage{csquotes}
\usepackage{comment}
\usepackage{bm}
\usepackage{units}
\usepackage{fancyhdr}
\usepackage{float}
\usepackage{capt-of}
\usepackage[bf, font=small,labelfont=bf]{caption}
\newcommand{\bcaption}[2]{\caption{\textbf{#1} #2}}
\usepackage{setspace}
\usepackage{alphalph}
\usepackage{mathtools}
\usepackage{authblk}
\usepackage[titletoc]{appendix}
\usepackage{amsmath}
\usepackage[backend=biber,style=numeric,sorting=none,maxbibnames=99,natbib]{biblatex}
\bibliography{main}
\usepackage[font=footnotesize]{subfig}

\usepackage{etoolbox}

\robustify{\subref}

\newcommand\numberthis{\addtocounter{equation}{1}\tag{\theequation}}
\title{\vspace{-2cm} A Novel Compartmental Approach to Modeling COVID-19 Disease Dynamics and Analyzing the Effect of Common Preventative Measures}
\author[1]{Caden Lin }
\affil[1]{ {\small The Harker School, San Jose, CA, 95120}}
\date{Revised November 2021}
\begin{document}
\setlength{\abovedisplayskip}{3pt}
\setlength{\belowdisplayskip}{3pt}
\begin{titlepage}
\clearpage\maketitle
\thispagestyle{empty}

\end{titlepage}

\thispagestyle{empty} 
\begin{abstract}
   As of December 2020, the COVID-19 pandemic has infected over 75 million people, making it the deadliest pandemic in modern history. This study develops a novel compartmental epidemiological model specific to the SARS-CoV-2 virus and analyzes the effect of common preventative measures such as testing, quarantine, social distancing, and vaccination. By accounting for the most prevalent interventions that have been enacted to minimize the spread of the virus, the model establishes a paramount foundation for future mathematical modeling of COVID-19 and other modern pandemics. Specifically, the model expands on the classic SIR model and introduces separate compartments for individuals who are in the incubation period, asymptomatic, tested-positive, quarantined, vaccinated, or deceased. It also accounts for variable infection, testing, and death rates. I first analyze the outbreak in Santa Clara County, California, and later generalize the findings. The results show that---although all preventative measures reduce the spread of COVID-19---quarantine and social distancing mandates reduce the infection rate and subsequently are the most effective policies, followed by vaccine distribution and, finally, public testing. Thus, governments should concentrate resources on enforcing quarantine and social distancing policies. In addition, I find mathematical proof that the relatively high asymptomatic rate and long incubation period are driving factors of COVID-19’s rapid spread. 
   
\end{abstract}
\pagebreak 
\setcounter{page}{1}
\section{Introduction}

\subsection{The COVID-19 Pandemic}
The emergence of a new strain of coronavirus named SARS-CoV-2 in Wuhan, China in December of 2019 has led to the global COVID-19 pandemic that has affected all worldwide operations. The respiratory infection---which causes symptoms such as a fever, cough, and difficulty breathing---has been confirmed to have infected over 75 million individuals and caused nearly 2.0 million deaths globally as of December 2020, though the actual numbers are significantly higher due to the presence of undetected infections \citep{symptoms, totalcases}. The casualty toll of the pandemic is far greater than any other disease outbreak in recent history and continues to grow worse \citep{MERS}. 


Unsurprisingly, many researchers have attempted to model COVID-19 \citep{CDCModels}. Compartmental epidemiological models are typically the most common, but many other methodologies are used \citep{mostcommon}. For instance, \citet{example_bayesian} use a Bayesian model, \citet{example_ridge} use ridge regression, and \citet{example_machine} use machine learning. Of these studies, some derive models specific to SARS-CoV-2 and the COVID-19 pandemic: \citet{ndairou} create a compartmental model that accounts for hospitalizations and super-spreaders, while \citet{criticalCaseModel} develop a compartmental model that considers a variable infection rate and critical vs. noncritical infections. Such models attempt to forecast the pandemic more accurately by considering nuances of the SARS-CoV-2 virus that make it particularly infectious.


In attempt to control the spread of the virus, governments around the world have imposed  preventative measures such as quarantines, shelter-in-place lockdowns, mask enforcement, and widespread testing \citep{govresponse}. Despite such attempts, cases around the world continue to rise, and the economic and social consequences of the pandemic grow worse. Governments also have limited resources, and citizens have a finite tolerance for preventative measures \citep{SocialandEcon, Pl20}. Still, such measures have been found to significantly reduce spread \citep{quarantine1, quarantine2, masks}. Thus, in addition to forecasting the virus's spread, identifying and analyzing common preventative measures are key to ending the pandemic. 

\subsection{Research Focus}

To my knowledge, none of the models specific to COVID-19 analyze the impact of individual preventative measures. In addition, none conduct quantitative investigations that study which measures are more effective than others. Because there is a finite amount of regulation that can be used enforced, it is important to study which measures should be prioritized. Therefore, this paper consists of three main goals: 

\begin{enumerate}
    \item Develop an accurate model specific to the COVID-19 pandemic that accounts for nuances of SARS-CoV-2 and common preventative measures implemented during the pandemic 
    \item Perform a case study with the model in one specific region to study its validity 
    \item Analyze the effect of common preventative measures and determine which are most effective 
\end{enumerate}

\section{Developing a Model}


\subsection{Methodology and Considerations}
Mathematical modeling can be classified as either data-driven---which allows one to derive equations directly from empirical data---or mechanistic, which uses the data to derive equations that can model a phenomenon. While data-driven models can make accurate predictions, they often fail to provide meaningful physical interpretations of the occurrence and only produce results. In other words, they can predict what will happen without explaining why. In this study, I make some assumptions that allow us to adopt mechanistic models for the research and derive a system of equations that can capture key dynamics of the outbreak such as the number of infected individuals for any amount of time in the future. The equations account for social distancing precautions, asymptomatic cases, testing, quarantine, a potential vaccine, and death. Throughout the study, social distancing will refer to any measure taken by all individuals to reduce the spread of the pandemic, which includes practices such as limiting non-essential travel or wearing masks. 

\begin{figure}[H]
    \centering
    {{ \includegraphics[width=14cm,height=7cm]{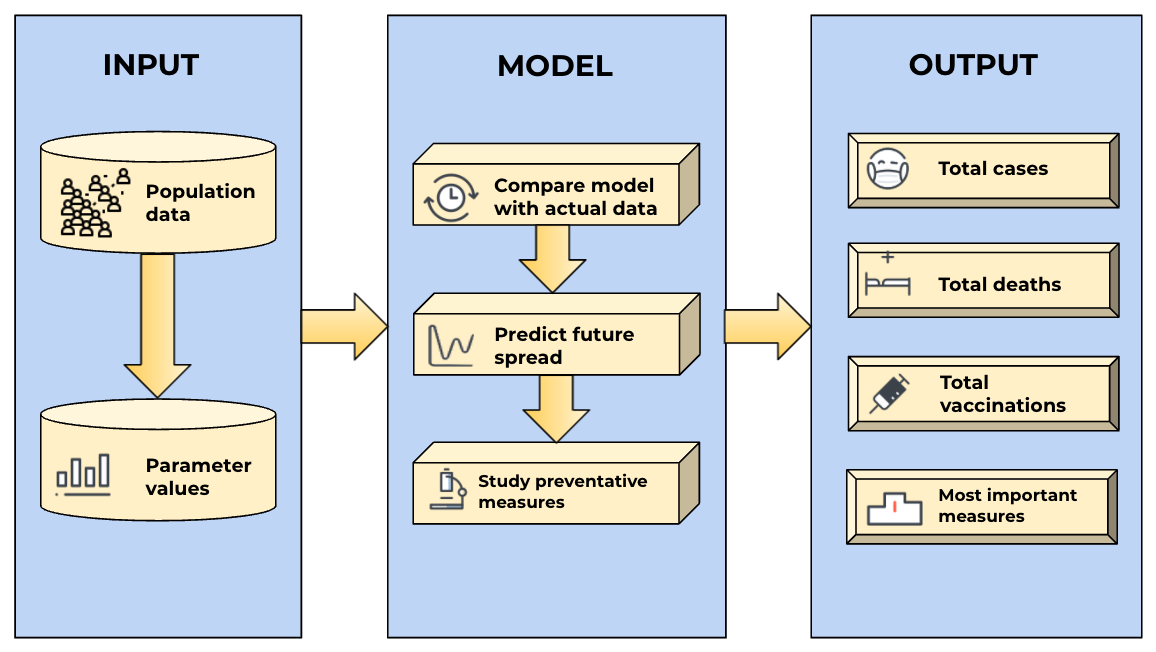} }}%
    \caption{\textbf{Overview of the Research Methodology and Goal}}
    \label{fig:nice diagram}
\end{figure}

\subsection{Developing the Model}
The most reasonable methodology is the utilization of a compartmental epidemiological model, a method first developed in 1927 as the result of research by mathematicians and epidemiologists Anderson Gray McKendrick and William Ogilvy Kermack \citep{SIR}. Compartmental models in epidemiology divide a given population into various categories and characterize the dynamics of each population using differential equations. For instance, the works of Kermack and McKendrick began with the simple SIR model, which creates three compartments: susceptible, infected, and recovered.

\begin{figure}[H]
    \centering
    {{ \includegraphics[width=7cm,height=3.6cm]{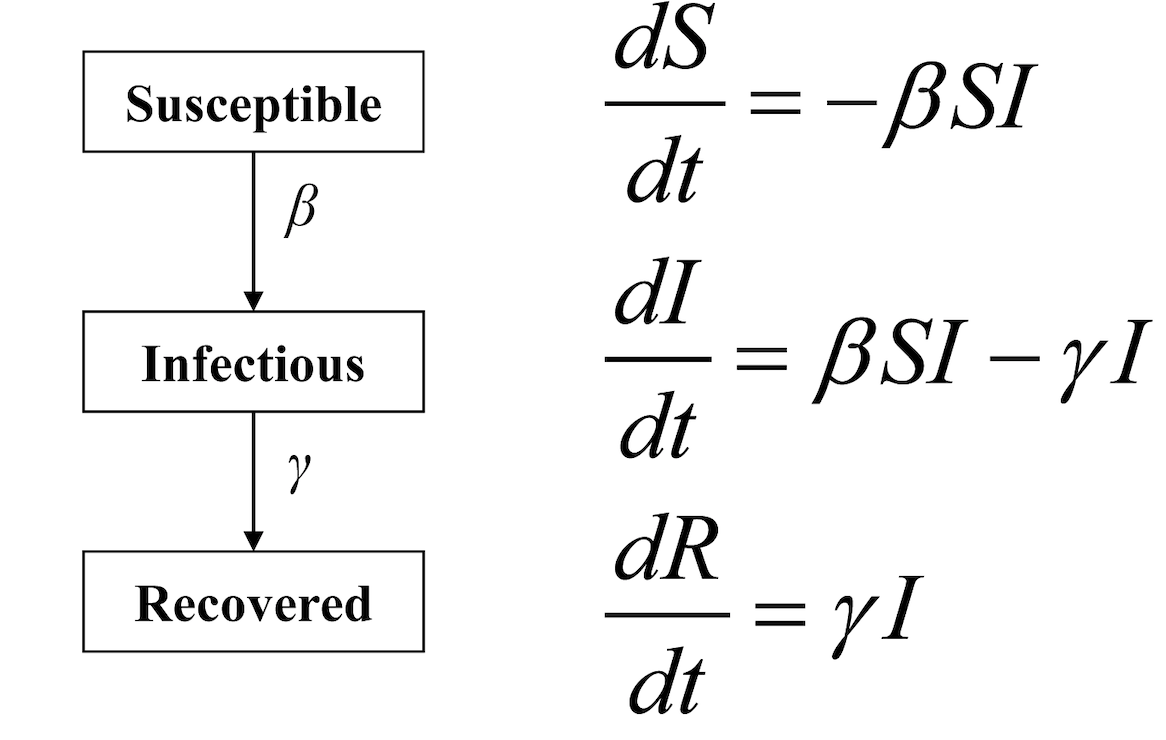} }}%
    \quad
    {{\includegraphics[width=7cm,height=3.6cm]{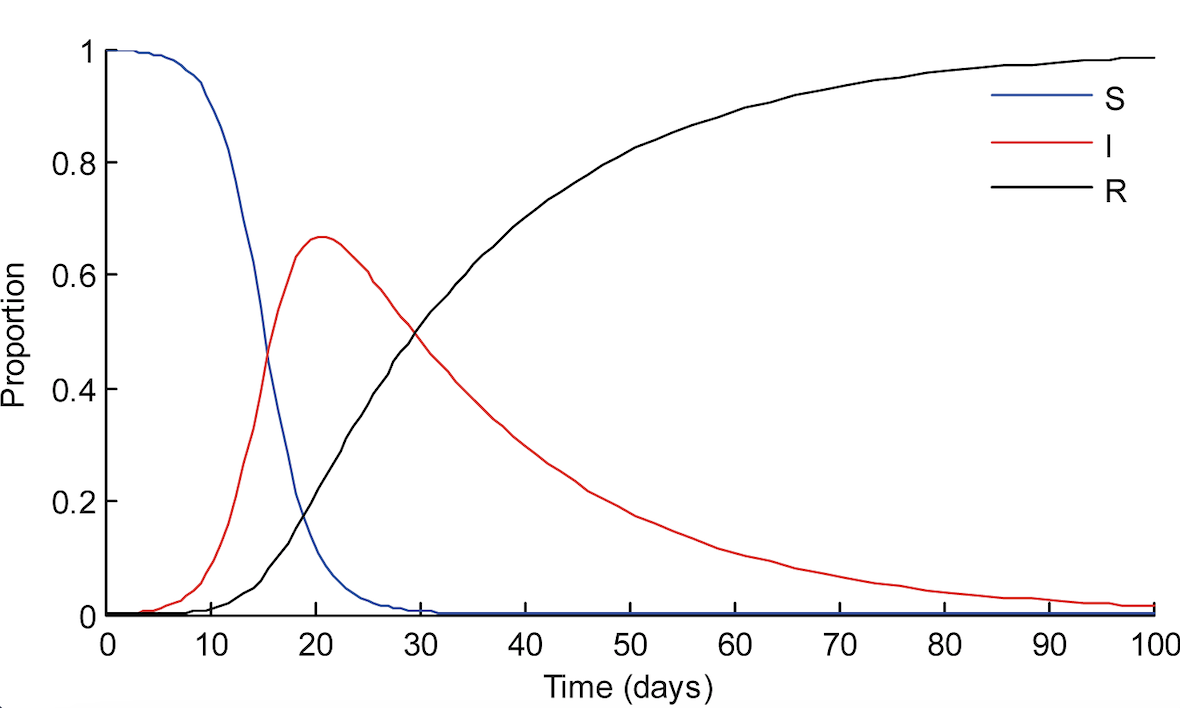} }}%
    \caption{\textbf{A visual representation of the SIR model and its equations \citep{sirmodeldiagram}}}
    \label{fig:SIR_Model}
\end{figure}

However, the model has since been adapted and expanded upon, leading to the creation of new compartmental models, such as the SEIR model---which considers incubation periods---and the MSIR model which accounts for maternally-derived immunity \citep{SEIR, MSIR}. Some compartmental models also account for variable contact rates, which are more realistic as they reflect changes in the infection rate as an outbreak progresses \citep{variable}. For this study, the following information should be considered:

\begin{enumerate}
    \item Research indicates that recovery from SARS-CoV-2 provides immunity \citep{immunity, immunity1}. This points to the need for a model with a designated compartment for recovered (i.e. immune) individuals. 
    \item Research  shows that a significant number of COVID-19 cases are asymptomatic \citep{asymptomatic1,asymptomatic2}. This results in a higher infection rate because asymptomatic individuals will not change their everyday behavior after being infected \citep{higherinfection}. Thus, I classify infections as either symptomatic or asymptomatic. 
    \item  SARS-CoV-2 has a notable incubation period where infected individuals do not exhibit symptoms but can spread the virus, pointing to the need for an incubation category \citep{incubation1, incubation2}. 
    \item There are a significant number of deaths resulting from COVID-19, pointing to the need for a specific category for deceased individuals \citep{deathrate}.  
    \begin{itemize}
        \item However, the death rate is not fixed. As hospitals and other healthcare services adapt to the pandemic, fewer infections are fatal. This suggests a variable death rate. 
    \end{itemize}
    \item There should be a compartment for vaccinated individuals who, like the recovered category, have immunity against the disease. The model should also consider the possibility that the vaccine is partially ineffective.
    \item I must account for variations in social distancing that would phenomenologically affect the infection rate of the disease, suggesting a variable infection rate. 
    \item Widespread testing can reduce the spread of COVID-19 by identifying asymptomatic individuals \citep{testing_reduces_covid}. However, the entire population cannot be tested. Thus, I classify asymptomatic individuals as having been tested or not. 
    \begin{itemize}
        \item However, as the availability of COVID-19 tests increases over time, more asymptomatic cases are identified, suggesting the need for a variable testing rate. 
    \end{itemize}
\end{enumerate}

\subsection{Outline of the Model}

Thus, the classifications that the model requires are as follows: susceptible ($S$), incubation ($I$), asymptomatic ($A$), asymptomatic-and-quarantined (due to a positive test, $P$), symptomatic-and-quarantined ($Q$), vaccinated ($V$), recovered ($R$), and dead ($D$). Also, the infection, testing, and death rates are variable, producing a SIAPQVRD model with variable rates. 

\begin{figure}[H]
    \centering
    \includegraphics[width=13cm, height=5.5cm]{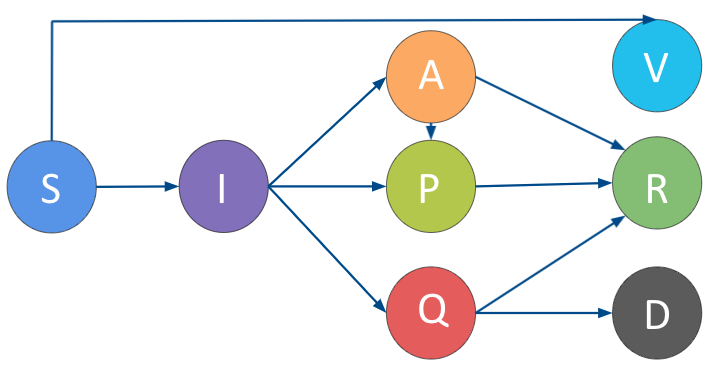}
    \caption{\textbf{A visual representation of the model}}
    \label{fig:visual}
\end{figure}

Susceptible individuals, who are the vast majority of the population, can become infected or vaccinated. If vaccinated successfully, an individual moves directly into the vaccinated category. If infected, an individual first enters an incubation period. Those who eventually exhibit symptoms move into the symptomatic (quarantined) category, while asymptomatic cases move to the asymptomatic compartment. 

Asymptomatic individuals ($I$ or $A$) who are tested for COVID-19 move into the tested-positive category. The model separates symptomatic (quarantined) individuals and asymptomatic individuals who have tested positive (also quarantined) because only symptomatic individuals can die of COVID-19. 

If symptomatic, individuals can  either move into the recovered category or the dead category. All asymptomatic cases recover.  

\subsection{Key Assumptions and Simplifications}

The following assumptions and simplifications are made and referenced  throughout the report. Justifications, explanations, and implications of the assumptions are discussed in the conclusion.

\begin{enumerate}
    \item The law of mass action is maintained. That is, the total rate of encounters (collision frequency) between individuals of two different categories is proportional to the product of their concentrations.
    \item Natural birth and death rates are negligible relative to the total population. 
    \item All individuals follow social distancing measures, and all individuals with a confirmed infection quarantine. 
    \item Only the vaccination of susceptible individuals is tracked.
    \item Only one dose of the vaccine is required, and, if successful, provides immunity immediately. 
    \item Only the testing of susceptible and asymptomatic individuals is tracked. 
    \item Testing is perfect (no false positives or negatives) and completely random.
\end{enumerate}

\section{Deriving the Model}
As shown in Figure \ref{fig:SIR_Model} and Figure \ref{fig:visual}, in compartmental epidemiological modeling, events such as infection or vaccination cause individuals to move from one category into another. In this section, I determine the mathematical basis of these movements. 

\subsection{Susceptible}

All individuals, except patient zero, begin susceptible. Traditionally, individuals can only move out of the susceptible population through infection; in the SIAPQVRD model, they can be vaccinated as well. Here, I will discuss the former, and vaccination will be incorporated in Section 3.6. 

The rate at which people become infected is controlled by the number of susceptible people, the number of infected people, and the infection rate. I incorporate a variable infection rate $\beta_t$ with the form:





\begin{center}
\begin{equation*}
 \beta_t = \begin{cases} 
     \beta_{t1} & 0\leq t\leq t_1 \\
     \beta_{t2} & t_1 < t\leq t_2 \\
     \beta_{t3} & t_2 < t\leq t_3 \\
     \vdots \\
     \beta_{tn} & t_{n-1} < t\leq t_n \\
   \end{cases} 
   \end{equation*}
   \end{center}
  
Each $\beta_{ti}$ represents a different stage of the infection rate. My model also considers unknown cases, which have different infection rates. Thus, I divide $\beta_t$ into two categories: $\beta_{\xi t}$ and $\beta_{\chi t}$, which refer to the infection rate for known and unknown cases respectively. Both will be divided as by constants, $\xi$ and $\chi$, to scale their values: 

\begin{equation*}
    \beta_{\xi t}=\frac{\beta_t}{\xi} \hspace{4cm} \beta_{\chi t}=\frac{\beta_t}{\chi}
\end{equation*} 

Since the SIAPQVRD has four categories of infected individuals that can all cause secondary infections, I must include all of them in the equation for $\frac{\mathrm{d}S}{\mathrm{d}t}$. Thus, 




\begin{equation}
\begin{split}
    \frac{\mathrm{d}S}{\mathrm{d}t} & = - \beta_{t} * S * (I+A+P+Q) - \mbox{[rate at which people are vaccinated]}\\
    & = - (\beta_{\xi t} * S (I+A) + \beta_{\chi t} * S(P+Q) ) - \mbox {[rate at which people are vaccinated]}
\end{split}
\end{equation}

\subsection{Incubation}

All infected individuals first enter an incubation period. To exit $I$, one must either exit the incubation period or be tested for COVID-19. Thus, 

\begin{equation*}
\begin{split}
     \frac{\mathrm{d}I}{\mathrm{d}t} & = \text{[rate of infection]} - \text{[rate at which people leave incubation]} - \text{[rate of testing]} \\ 
    & = (\beta_{\xi t} * S (I+A) + \beta_{\chi t} * S (P+Q) ) - \text{[rate at which people leave incubation]} - \text{[rate of testing]}
\end{split} 
\end{equation*}

Let $\lambda$ be the incubation rate, or the fraction of the population $I$ that exits the incubation period daily. Let $\omega_t$ be the variable testing rate: a fraction $\omega$ of the tested population is tested daily. Let $\epsilon$ be the testing parameter. I incorporate a variable testing rate:

\begin{center}
\begin{equation*}
\omega_t = \begin{cases} 
     \omega_{t1} & 0\leq t\leq T_1 \\
     \omega_{t2} & T_1 \leq t\leq T_2 \\
     \omega_{t3} & T_2 \leq t\leq T_3 \\
     \vdots \\
     \omega_{ti} & T_{i-1} \leq t\leq T_i \\
   \end{cases} 
\end{equation*}
\end{center}
The testing parameter $\epsilon$ scales the testing rate by a constant, which allows us to tune the testing rate during my analysis. For instance, doubling $\epsilon$ and simulating an outbreak under otherwise equal conditions explores the effect of doubling the testing rate. Thus,

\begin{equation}
    \begin{split}
         \frac{\mathrm{d}I}{\mathrm{d}t} & = (\beta_{\xi t} * S (I+A) + \beta_{\chi t} * S *(P+Q) ) - \lambda I - \omega_t \epsilon I \\
        & =  \frac{\beta_t}{\xi}*S(I+A) +  \frac{\beta_t}{\chi}*S(P +Q) - \lambda I - \omega_t \epsilon
    \end{split}
\end{equation}

\subsection{Asymptomatic}

Individuals enter the asymptomatic category from the incubation category and exit to the recovered category or the tested-positive category. Thus, 

\begin{equation*}
    \begin{split}
     \frac{\mathrm{d}A}{\mathrm{d}t} & = \text{[rate at which asymptomatic cases leave incubation period]} \\
    & - \text{[rate at which people are tested]} -\text{[rate at which people recover]}
    \end{split} 
\end{equation*}

Let $\alpha$ be the rate of asymptomatic infections, or the fraction of total cases that are asymptomatic. Let $ \gamma_A$ be the recovery rate for asymptomatic infections: a fraction $\gamma_A$ of the asymptomatic category recovers each day. Thus,

\begin{equation}
     \frac{\mathrm{d}A}{\mathrm{d}t} = \alpha \lambda I - \omega_t \epsilon A - \gamma_A A 
\end{equation}

\subsection{Tested-Positive}

People move into $P$ from $I$ or $A$ by being tested. The only way to exit $P$ is by recovering. Thus:
\begin{equation*}
     \frac{\mathrm{d}P}{\mathrm{d}t} = \text{[rate at which people are tested]} - \text{[rate at which people recover]} 
\end{equation*}

Anyone in $P$ is asymptomatic and therefore has a recovery rate $\gamma_A$. Thus, 

\begin{equation}
     \frac{\mathrm{d}P}{\mathrm{d}t} = \omega_t \epsilon I +  \omega_t \epsilon A - \gamma_A P
\end{equation}
\subsection{Quarantine}

People enter $Q$ by developing a symptomatic infection and consequently quarantining. People exit $Q$ by recovering or dying. Thus, 
\begin{equation*}
    \begin{split}
     \frac{\mathrm{d}Q}{\mathrm{d}t} = &  \text{[rate at which symptomatic infections emerge]} \\
        &-\text{[rate of recovery (symptomatic)]} - \text{[rate of death]}
    \end{split} 
\end{equation*}

Any infection that is not asymptomatic is symptomatic, so the rate of symptomatic infections is $1-\alpha$. Let $\gamma_S$ be the recovery rate for symptomatic infections and let $\delta_t$ be the variable death rate (a fraction $\delta$ of $Q$ dies daily). 

\subsubsection{Death Rate}

The death rate changes over time as hospitals and other services adapt to the pandemic. Thus, I include a variable death rate: 

\begin{center}
\begin{equation*}
 \delta_t = \begin{cases} 
     \delta_{t1} & 0\leq t\leq \tau_1 \\
     \delta_{t2} & \tau_1 \leq t\leq \tau_2 \\
     \delta_{t3} & \tau_2 \leq t\leq \tau_3 \\
     \vdots \\
     \delta_{tk} & \tau_{k-1} \leq t\leq \tau_k \\
   \end{cases} 
    \end{equation*}
   \end{center}  

Thus,
\begin{equation}
 \frac{\mathrm{d}Q}{\mathrm{d}t} = (1-\alpha) \lambda I - \gamma_s Q - \delta_t Q 
\end{equation}

\subsection{Vaccinated}

People enter $V$ from $S$ by receiving a successful vaccination, and people cannot exit $V$. Let $\theta$ be the vaccination rate (a fraction $\theta$ of $S$ is vaccinated daily). Let $\rho$ be the success rate of the vaccination (a fraction $\rho$ of vaccines are successful in preventing COVID-19). Unsuccessful vaccines have no effect. Combined, a product $\theta$ and $\rho$ of $S$ are vaccinated daily. Therefore, 

\begin{equation}
     \frac{\mathrm{d}V}{\mathrm{d}t} = \theta \rho S 
\end{equation}

From equation (1), we now have:
\begin{equation}
\begin{split}
    \frac{\mathrm{d}S}{\mathrm{d}t} & = - (\beta_{\xi t} * S (I+A) + \beta_{\chi t} * S(P+Q) ) - \theta \rho S \\
    & = - \frac{\beta_t}{\xi}*S(I+A)  - \frac{\beta_t}{\chi}*S(P+Q) - \theta \rho S
\end{split}
\end{equation}

\subsection{Recovered}

People enter $R$ by recovering from an infection, either symptomatic or asymptomatic, and cannot exit: 

\begin{equation}
     \frac{\mathrm{d}R}{\mathrm{d}t} = \gamma_A (A + P) + \gamma_S Q 
\end{equation}

\subsection{Dead}

People enter $D$ by dying and cannot exit: 

\begin{equation}
     \frac{\mathrm{d}D}{\mathrm{d}t} = \delta_t Q 
\end{equation}

\subsection{Complete System} 
The complete system is: 

\begin{equation}
\begin{aligned}
    \frac{\mathrm{d}S}{\mathrm{d}t} &= - \frac{\beta_t}{\xi}*S(I+A)  - \frac{\beta_t}{\chi}*S(I+A) -\theta \rho S \\ 
    \frac{\mathrm{d}I}{\mathrm{d}t} &=  \frac{\beta_t}{\xi}*S(I+A) +  \frac{\beta_t}{\chi}*S(I+A) - \lambda I - \omega_t \epsilon  I \\
    \frac{\mathrm{d}A}{\mathrm{d}t}&= \alpha \lambda I -  \omega_t \epsilon A - \gamma_A A \\
    \frac{\mathrm{d}P}{\mathrm{d}t}&= \omega_t \epsilon I +  \omega_t \epsilon A - \gamma_A P \\
    \frac{\mathrm{d}Q}{\mathrm{d}t}&=  (1-\alpha)\lambda I - \gamma_S Q - \delta_t Q \\
    \frac{\mathrm{d}V}{\mathrm{d}t}&=\theta \rho S \\
    \frac{\mathrm{d}R}{\mathrm{d}t}&=\gamma_A(A+P) + \gamma_S Q  \\
    \frac{\mathrm{d}D}{\mathrm{d}t}&=\delta_t Q \\
    \label{total_system}
\end{aligned}
\end{equation}

\vspace{-0.5cm}
\begin{figure}[H]
    \centering
    \includegraphics[width=13cm, height=6cm]{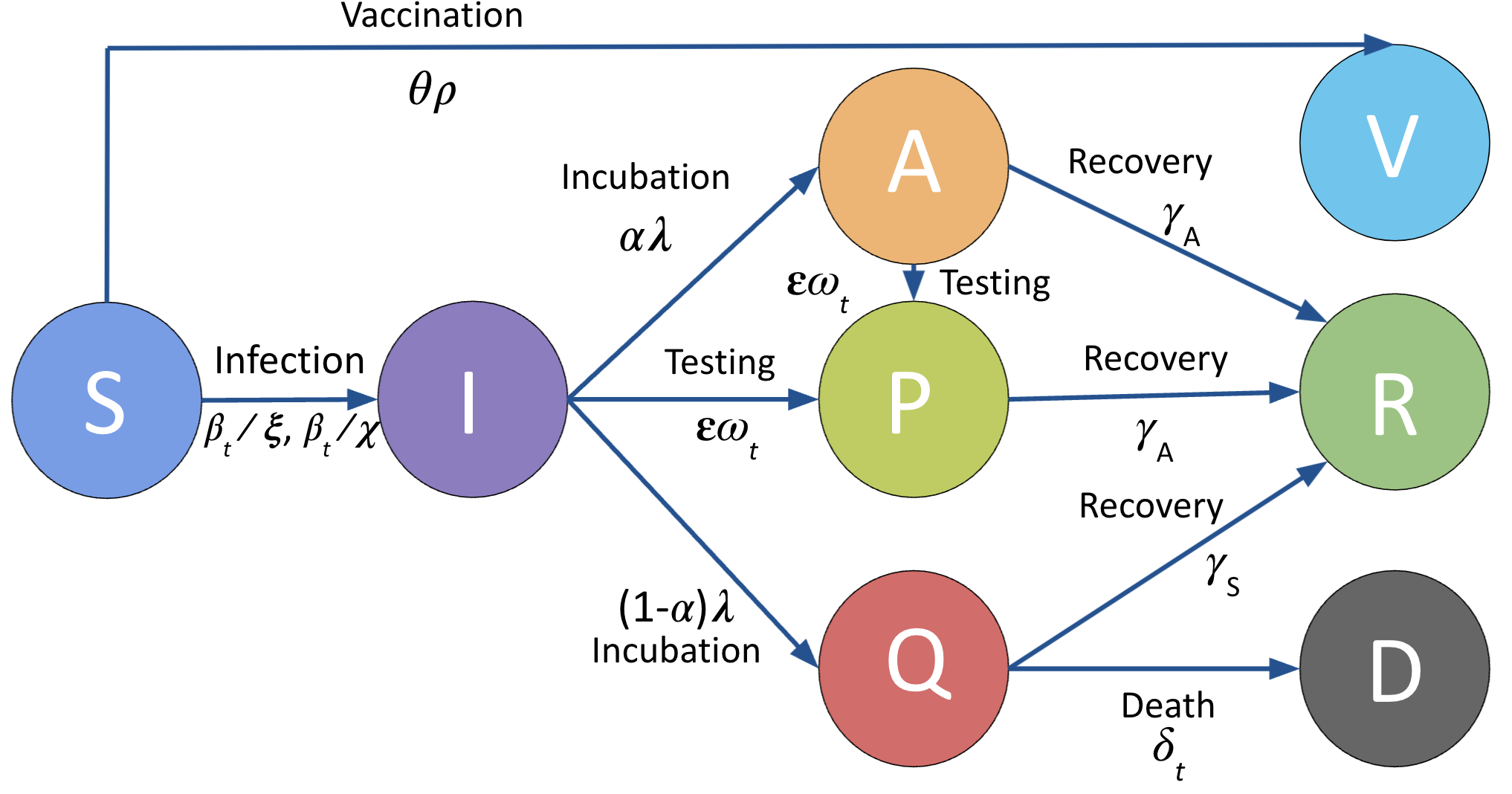}
    \caption{\textbf{A visual representation of the model with rates of population flow}}
\end{figure}

\noindent The total population $N$ should not change. That is, 
\begin{align*}
    N &= S+I+A+P+Q+V+R+D \\
    \Rightarrow \frac{\mathrm{d}N}{\mathrm{d}t} &= \frac{\mathrm{d}S}{\mathrm{d}t} +  \frac{\mathrm{d}I}{\mathrm{d}t} +  \frac{\mathrm{d}A}{\mathrm{d}t} +  \frac{\mathrm{d}P}{\mathrm{d}t} +  \frac{\mathrm{d}Q}{\mathrm{d}t} +  \frac{\mathrm{d}V}{\mathrm{d}t} +  \frac{\mathrm{d}R}{\mathrm{d}t} +  \frac{\mathrm{d}D}{\mathrm{d}t} = 0  \numberthis \label{eqn}
\end{align*}

\noindent This can be confirmed by taking the sum of the equations in (\ref{total_system}). 

Like most compartmental epidemiological models, each compartment will represent a fraction of the total population, rather than the true number of individuals in the category. Thus, $N=1$. 
\\
\subsection{Summary of State Variables and Parameters}

\indent \indent $S(t) = $ the fraction of the population that is susceptible. 

$I(t) = $ the fraction of the population that is in the incubation period.

$A(t) = $ the fraction of the population that is asymptomatic and has not been tested and is therefore not quarantined.  

$P(t) = $ the fraction of the population that is asymptomatic and has been tested and quarantined. 

$Q(t) = $ the fraction of the population that is quarantined because they are exhibiting symptoms.

$V(t) = $ the fraction of the population that is vaccinated. 

$R(t) = $ the fraction of the population that has recovered.

$D(t) = $ the fraction of the population that has died of infection. 

\indent $\beta_{\xi t}$ = the variable infection rate for asymptomatic individuals that accounts for preventative measures that are taken by the entire population (e.g., social distancing or mask-wearing). $\xi$ controls the effectiveness of these measures. Each asymptomatic person infects, on average, $\beta_{\xi t}$ other people per day. 

$\beta_{\chi t} = $ the infection rate for individuals who are quarantined (symptomatic or asymptomatic and tested-positive), who therefore have a lower infection rate \citep{functionalfear}. $\chi$ governs how strict the quarantine is. Each symptomatic or tested-positive person infects, on average, $\beta_{\chi t}$ other people per day. 

$\gamma_A = $ the recovery rate for asymptomatic individuals. A fraction $\gamma_A$ of the asymptomatic population recovers every day. $\frac{1}{\gamma_A}$ is the average recovery duration for asymptomatic infections. 

$\gamma_S = $ the recovery rate for symptomatic individuals. A fraction $\gamma_S$ of the symptomatic population recovers every day. $\frac{1}{\gamma_S}$ is the average recovery duration for symptomatic infections. 

$\lambda$ = the rate at which people with symptomatic infections begin to exhibit symptoms. $\frac{1}{\lambda} = $ the average incubation period.

$\omega_t = $ the variable testing rate. A fraction $\omega_t$ of the population is tested daily. 

$\alpha = $ the asymptomatic rate. A fraction $\alpha$ of cases are asymptomatic. 

$\theta = $ the rate of vaccination, which reflects the availability of the vaccine. A fraction $\theta$ of the susceptible population is vaccinated every day. 

$\rho = $ the success rate of the vaccine. A fraction $\rho$ of vaccinations prevent infection.

$\delta_t$ = the variable death rate. A fraction $\delta_t$ of the symptomatic population dies daily.

\section{Determining Parameter Values}

To determine specific parameter values, I apply the model to Santa Clara County, CA, which has a population of 1,928,000 and 75,000 cumulative cases at the time of writing \citep{population, SCCwebsite}. 


\subsection{General Parameters}

Some parameters are inherent to the SARS-CoV-2 virus: 

\begin{center}
\begin{table}[H]
\centering
\begin{tabular}{ |c|c|c| } 
 \hline
 Parameter & Meaning & Value  \\ 
 \hline
 $\alpha$ & Asymptomatic rate & 0.31 \\ 
 \hline
  $\lambda$ & Incubation rate & $\nicefrac{1}{5.7}$ \\ 
 \hline
  $\gamma_A$ & Asymptomatic recovery rate & $\nicefrac{1}{19}$ \\ 
 \hline
  $\gamma_S$ & Symptomatic recovery rate & $\nicefrac{1}{14}$ \\ 
 \hline
  $\rho$ & Vaccine success rate & 0.90 \\ 
 \hline
\end{tabular}
\caption{ \textbf{General parameter values \cite{alpharate, incubationrate, recoveryrateasymp,recoveryratesymp,pfizer}} }
\end{table}
\end{center}

\vspace{-1cm}

\noindent The remaining parameters are calculated specifically for Santa Clara County. In the following calculations, data collection began on March 2, 2020 ($t=0$) and ended on December 7, 2020 ($t=280$). 

\subsection{Infection Rate}

To determine $\beta_t$, I use publicly released information on COVID-19 in Santa Clara County and calculate the infection rate by dividing the number of new cases per day by the total number of known, active cases. Then, I take a seven-day moving average of the values, yielding: 

\begin{figure}[H]
    \centering
    \includegraphics[width=12cm, height=5.5cm]{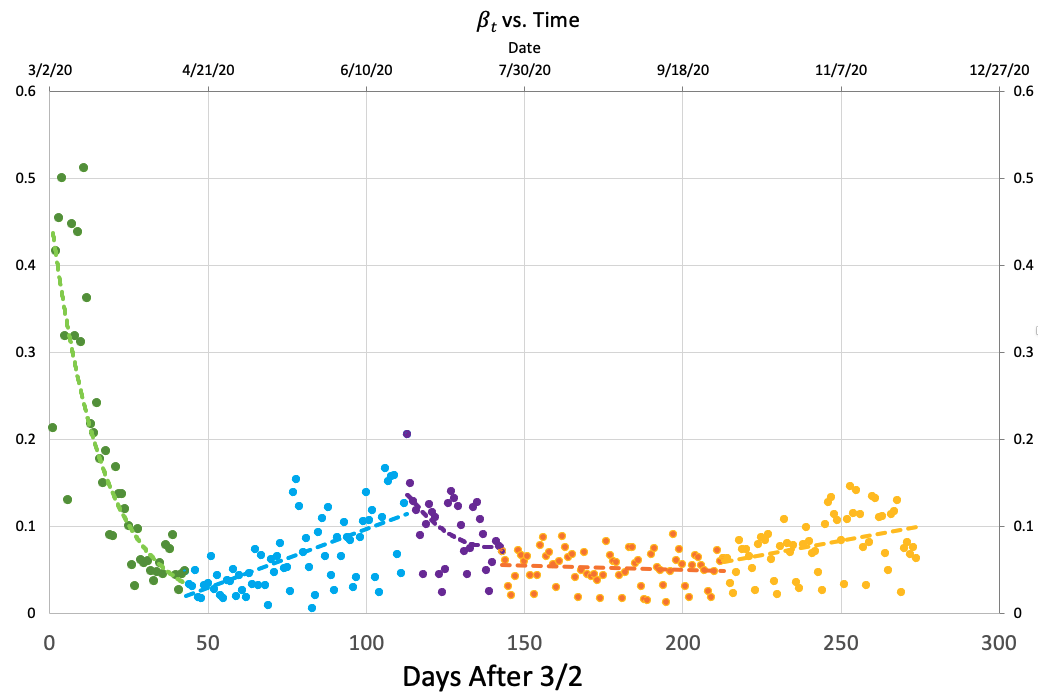}
    \caption{\textbf{The infection rate over time.} I distinguish five distinct stages of $\beta_t$. Infection rates have grown less volatile throughout of the pandemic, which makes intuitive sense as healthcare services have adapted to the pandemic over time.}
\end{figure}

\[ \beta_t = \begin{cases} 
     0.4645e^{-0.0610t} & 0\leq t\leq 43, \mbox{  March 2nd to April 14th} \\
      0.00135702t - 0.03865422 & 43< t\leq 113, \mbox{  April 15th to June 23rd}\\
      0.00008619t^2 - 0.02403308t + 1.75188688 & 113< t\leq 143, \mbox{  June 24th to July 23rd}\\
     -0.00009458t + 0.06901033 & 143< t\leq 213, \mbox{ July 24th to October 1st}\\
     0.00075163t - 0.10358623 & 213<t, \mbox{ October 2nd onwards }
   \end{cases} \]

\subsubsection{Approximation of Current Cases}

Because there is no way to determine the true number of active cases at any given time, I make an approximation. Letting $C_t$ represent the number of known active cases at time $t$:

\begin{equation}
    \begin{split}
        C_t & \approx \text{[known cases]}_t - \text{[known cases]}_{t - \text{average recovery time}} \\
        & = (A+P)_t - (A+P)_{t- (\alpha \gamma_A + (1-\alpha) \gamma_S)} \\
        & \approx (A+P)_t - (A+P)_{t- 17} 
    \end{split}
\end{equation}



\subsubsection{Accounting for Asymptomatic Infections}

In calculating $\beta_t$, I assume that the number of known cases reported by the county is equal to the total number of infections, as it is impossible to determine the true number of asymptomatic cases. Due to this approximation, this $\beta_t$ is not a true representation of the infection rate as it fails to account for asymptomatic cases. However, the function still provides a mechanistic view of the infection rate over time: if this approximated $\beta_t$ increases, then the true infection rate  also is increasing. I assume that the difference between the calculated $\beta_t$ and the true infection is the absence of a scalar multiple, which is taken into consideration by the parameters $\xi$ and $\chi$. I will estimate the specific values of $\xi$ and $\chi$ once the remaining parameters are determined.  

\subsection{Testing Rate}

I perform a similar analysis with the testing rate. I take the number of raw tests daily and divide it by the total population to determine $\omega_t$. 

\begin{figure}[H]
    \centering
    \includegraphics[width=12cm, height=5cm]{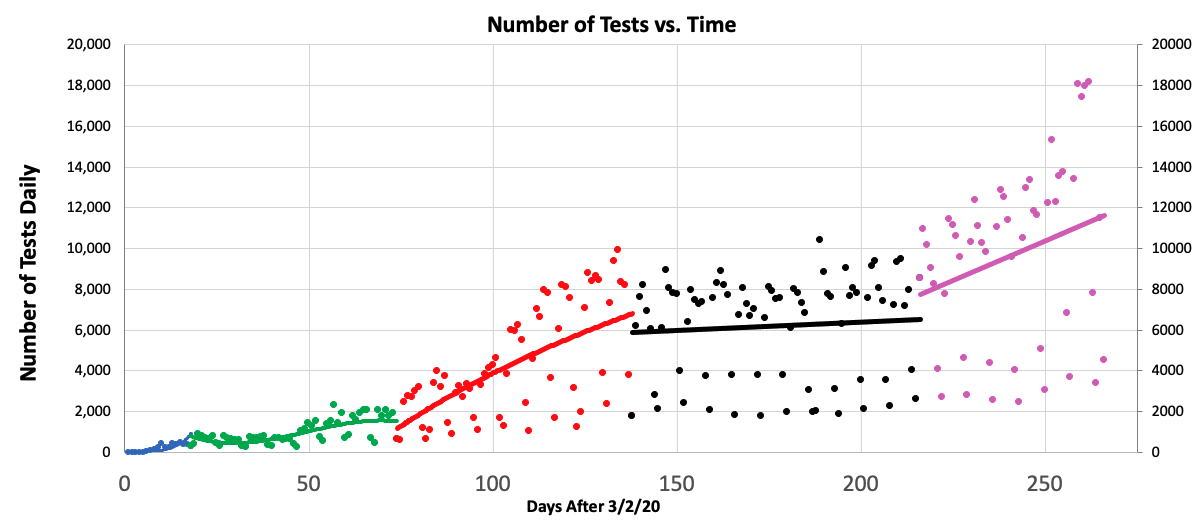} 
    \bcaption{The number of daily tests over time.} {Note that this graph shows the raw number of tests, rather than as a fraction of the total population. There is a second, lower sequence of data points because the county performs less testing on weekends.}
\end{figure}

  \[ \omega_t = \frac{1}{1928000} * \begin{cases} 
    9.0525e^{0.2574 t} & 0<t\leq19, \mbox{  March 2nd to March 20th} \\
   -0.0349t^3 + 5.162t^2 - 213.236t + 3156.424 & 19< t\leq 74, \mbox{  March 21st to May 15th}\\
     9069.32\ln{t} - 37864.8 & 75< t\leq 138, \mbox{  May 16th to July 18th}\\
    7.8545t + 4819.2 & 139< t\leq 216, \mbox{ July 19th to October 4th}\\
    76.876t - 8834.6 & 216<t, \mbox{ October 5th onwards}
   \end{cases} \]

\subsection{Death Rate}

The death rate is determined in the same way. I take the number of daily deaths and divide them by the estimated number of active cases to obtain the death rate $\delta$ on that day.

\begin{figure}[H]
    \centering
    \includegraphics[width=12cm, height=5.5cm]{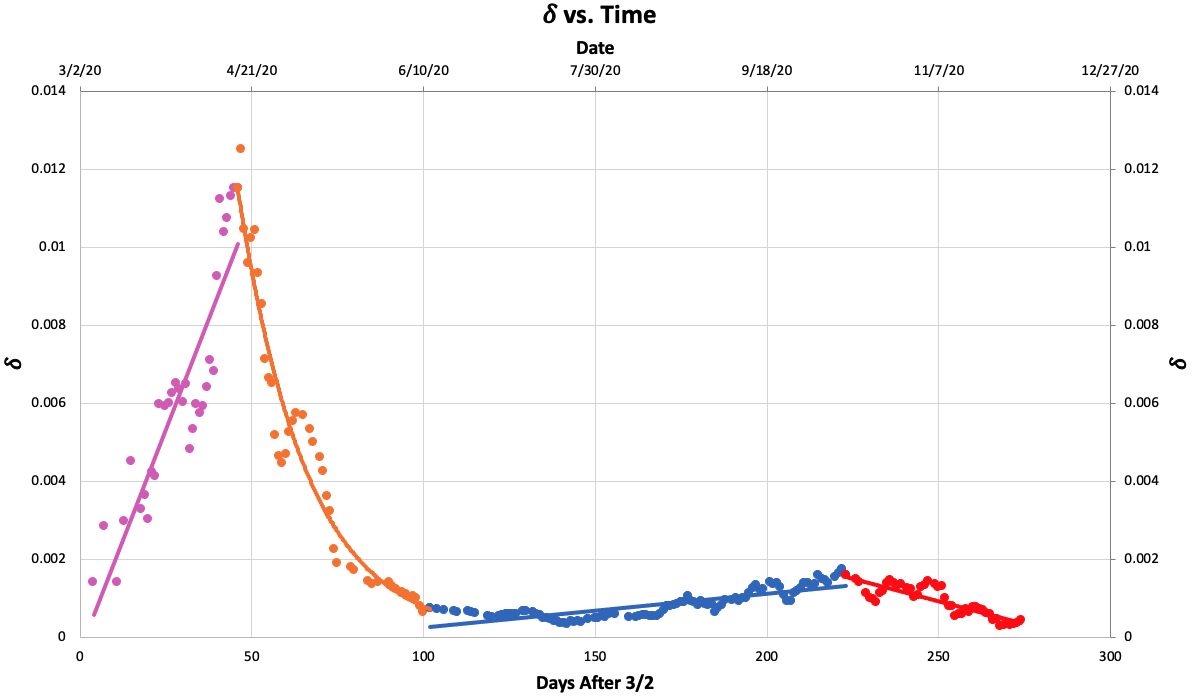}
    \bcaption{Death rate over time.}{I assume that all the reported cases have the potential to cause death; that is, every reported case is symptomatic. In reality, though, I cannot determine how many of the reported cases are symptomatic or asymptomatic. As I discover later, this assumption is almost insignificant. }
\end{figure}

This plot yields: 
   \[ \delta_t =  \begin{cases} 
   0.000227t - 0.000345 & t\leq 44, \mbox{  March 2nd to April 16th} \\
  0.11237e^{-0.04968t} & 44<t\leq 102, \mbox{  April 17th to June 12th} \\
  0.00000869t - 0.00062901 & 102 < t \leq 223, \mbox{  June 13th to October 11th}\\
  -0.00002283t + 0.00663369 & 223 < t, \mbox{October 12th onwards} 
   \end{cases} \]

\subsection{Determining $\xi$ and $\chi$}

Unlike other parameters, there is no official value for $\xi$ or $\chi$: I must determine it through experimentation. The role of $\xi$ and $\chi$ is twofold: first, to consider that people who know they are infected will quarantine themselves, while others will only follow common measures such as social distancing; and second, to account for the simplification made in determining $\beta_t$ that disregards asymptomatic cases. If these purposes are unaccounted for, then the infection rate used will be too large, as the the infection rate is calculated by dividing the daily number of new cases by the product of $S$ and the total cases. If I assume that all cases are reported, then I ignore asymptomatic cases, and the total cases will be lower than they should be. 



Since $\beta_t$ was calculated by disregarding asymptomatic cases, it reflects regular social behavior of people who do not believe that they are infected, regardless of whether they actually are. Thus, by default, I set $\xi = 1$. This approximation does not stand for individuals who know they are infected, as they would quarantine, thereby reducing their infection rate. To determine the exact value of $\chi$, I experiment with different values and simulate hypothetical outbreaks until I most closely resemble the actual outbreak in Santa Clara County, which occurs at $\chi = 1.75$. 



\subsection{Applying the Model to the COVID Outbreak in Santa Clara County}
I now have a complete model with parameter values and functions. To test the validity of the model, I assumed initial conditions based on the COVID-19 data for Santa Clara County.

\begin{center}
\begin{tabular}{ |c|c|c| } 
 \hline
 Population & Initial Value  \\ 
 \hline
 $S$ & 0.991141\\ 
 \hline
  $I$ & ~0.000001  \\ 
 \hline
  $A$ & ~0.000001 \\ 
 \hline
  $P$ & ~0.00000241182\\ 
  \hline
\end{tabular}
\qquad
\qquad
\qquad
\begin{tabular}{|c|c|c|}
 \hline
 Population & Initial Value  \\ 
 \hline
  $Q$ & ~0.00000241182 \\ 
 \hline
  $V$ & 0 \\ 
 \hline
  $R$ & 0.00884728 \\ 
 \hline
  $D$ & 0.000001556 \\ 
 \hline
\end{tabular}
\end{center}

Numerically solving the system using a Fourth-Order Runge-Kutta approximation in Mathematica returns: 

\begin{figure}[H]
    \centering
    \subfloat[\textbf{Predicted number of cases.} The sum of the green ($P$) and red ($Q$) curves should be equal to the blue curve, which represents the actual known cases. ]{{ \includegraphics[width=7.3cm,height=3.7cm]{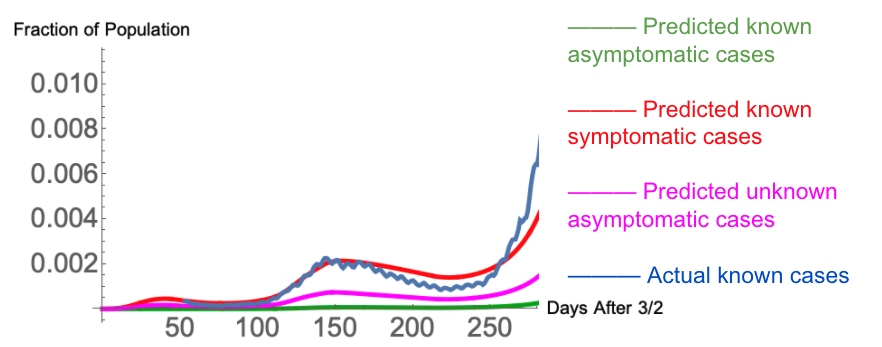} }}%
    \qquad
    \subfloat[\textbf{Predicted number of deaths}]
    {{\includegraphics[width=7.3cm,height=3.7cm]{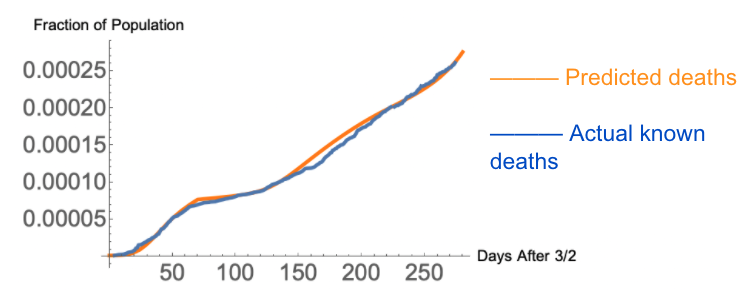} }}%
   \bcaption{Predicted outbreak vs. actual outbreak.}{Discrepancies in the model can be attributed to the assumptions and simplifications made and approximations in calculating the parameter values.}%
    \label{fig:predict and actual}%
\end{figure}

Figure \ref{fig:predict and actual}a shows that there is a significant population of asymptomatic cases ($A$, shown in magenta) at all times, demonstrating that the number of cases is drastically underreported. The value of $P$, shown in green, is very small, which indicates that testing does not identify a significant number of cases. This also explains why my assumption in calculating the death rate is almost insignificant: almost all of the reported cases are symptomatic and therefore could cause death. However, the effect of testing may be understated due to the assumption of random testing. In reality, testing is not random; those who believe they may have been exposed to the virus are more likely to be tested. 



\section{Analyzing Preventative Measures}

The most significant purpose of this study is to analyze the effect that preventative measures have on containing the pandemic and determine which measures are most significant. Because the model includes parameters that dictate different aspects of the virus or different preventative measures, parameters can be tuned to explore possible outbreaks under different conditions. To accomplish this, the independent variable must be a parameter instead of time; and, to isolate the effect of each parameter, only one parameter will be tuned at once. The dependent variable will be the number of cases and deaths by $t=280$. 

\subsection{Social Distancing and Quarantine}

Social distancing and quarantine are two of the most common preventative measures \citep{quarantine}. In the model, they are represented by $\xi$ and $\chi$ respectively and scale the infection rate. Thus, tuning $\xi$ and $\chi$ illustrates the significance of the infection rate:

\begin{figure}[H]
    \centering
    \subfloat[\textbf{Total cases and deaths vs. \boldmath$\xi$.} Increasing $\xi$ by 25\% (decreasing the infection rate for unknown cases by 20\%) to 1.25 would decrease the total number of cases and deaths by nearly twofold. Likewise, decreasing $\xi$ by just 10\% to 0.9 would increase the number of cases and deaths by approximately a factor of three. ]{{ \includegraphics[width=12cm,height=6cm]{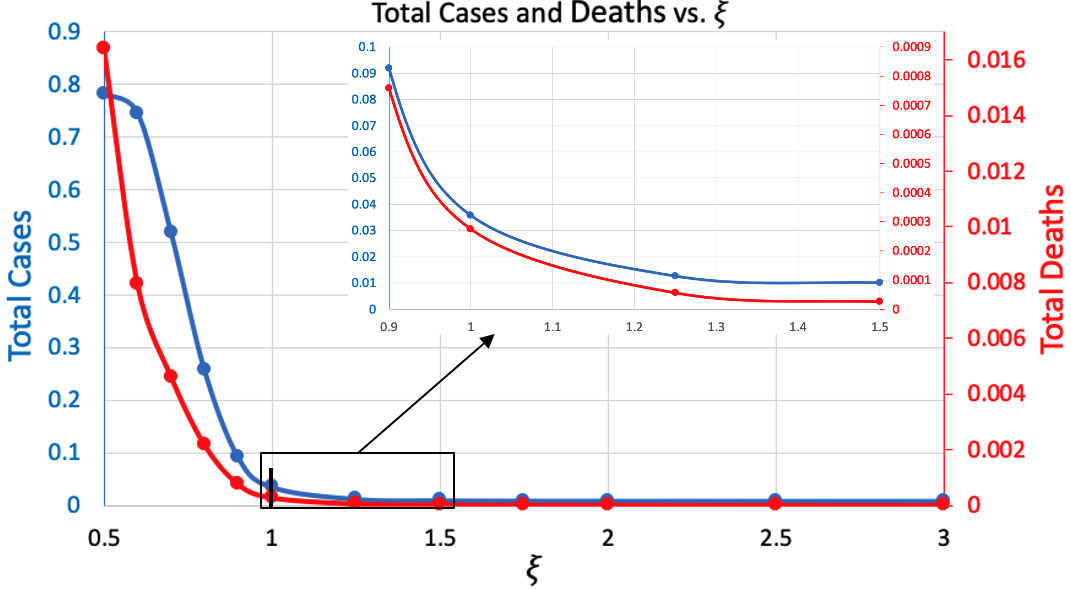} }}%
     \quad
    \subfloat[\textbf{Total cases and deaths vs. \boldmath$\chi$.} Increasing $\chi$ from 1.75 to 2 could reduce total cases and deaths by almost twofold. Similarly, a reduction in $\chi$ to 1.5 would more than double the number of total cases and deaths.]
    {{\includegraphics[width=12cm,height=6cm]{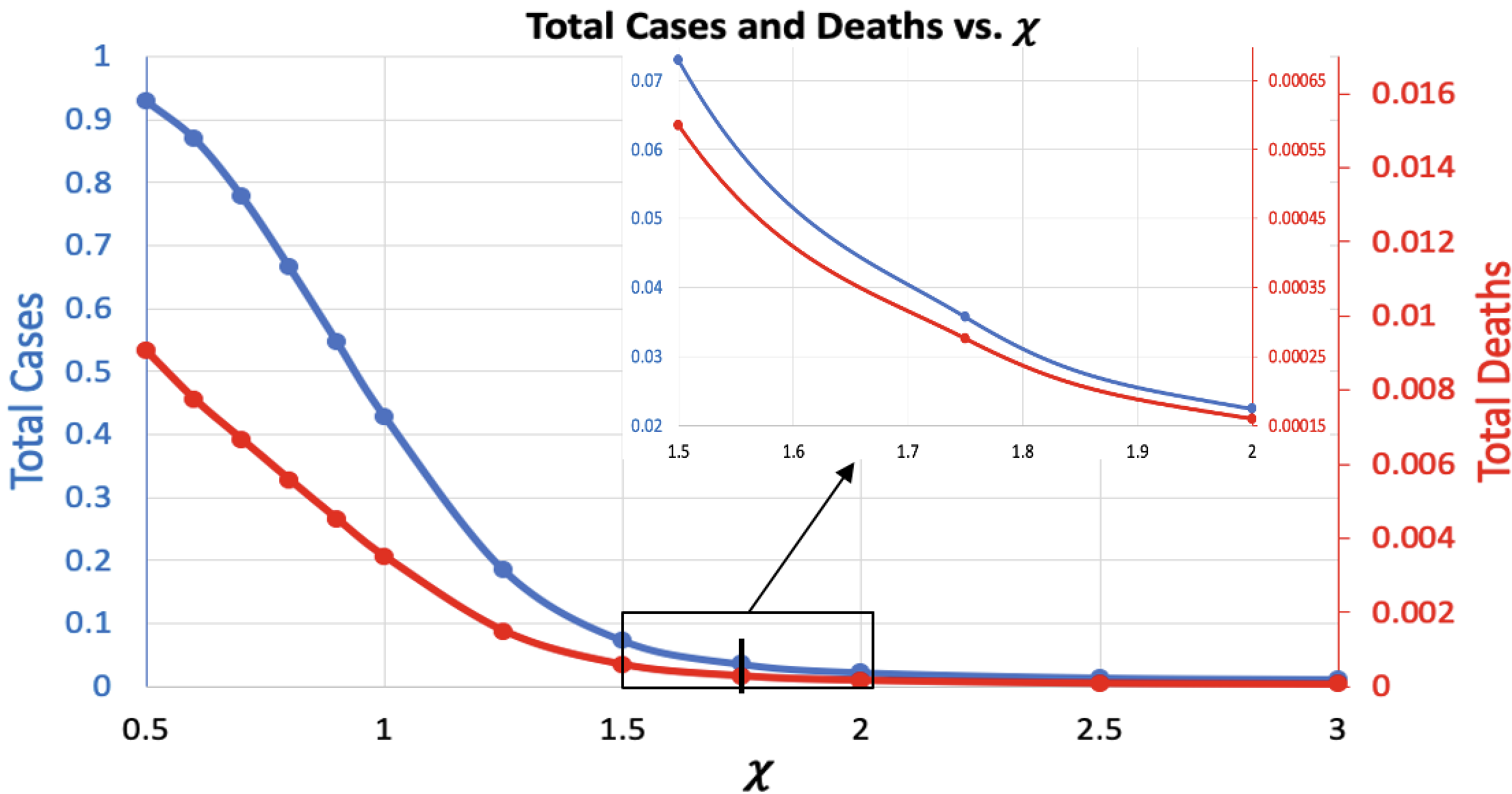} }}%
   \caption{\textbf{Total cases and deaths at different $\xi$ and $\chi$ }}%
    \label{fig:xi and chi}%
\end{figure}

Figure \ref{fig:xi and chi} shows that $\xi$ and $\chi$ have a very significant effect on the dynamics of the outbreak. The relationship is exponential: the infection rate has a significantly larger effect at larger values (smaller values of $\xi$ and $\chi$). At smaller infection rates, variations in $\xi$ and $\chi$ have little to no effect, as demonstrated up close for $1.3 < \xi < 1.5$ in \ref{fig:xi and chi}a. 

\subsection{Testing}

Testing has been adopted and strongly encouraged strongly by governments to identify asymptomatic cases and track the outbreak \cite{testtesttest}. More testing slows the spread of the virus because more asymptomatic cases are identified and more quarantining occurs. With the addition of $\epsilon$, I can tune the testing rate $\omega_t$. 

\begin{figure}[H]
    \centering
    \includegraphics[width=12cm,height=6cm]{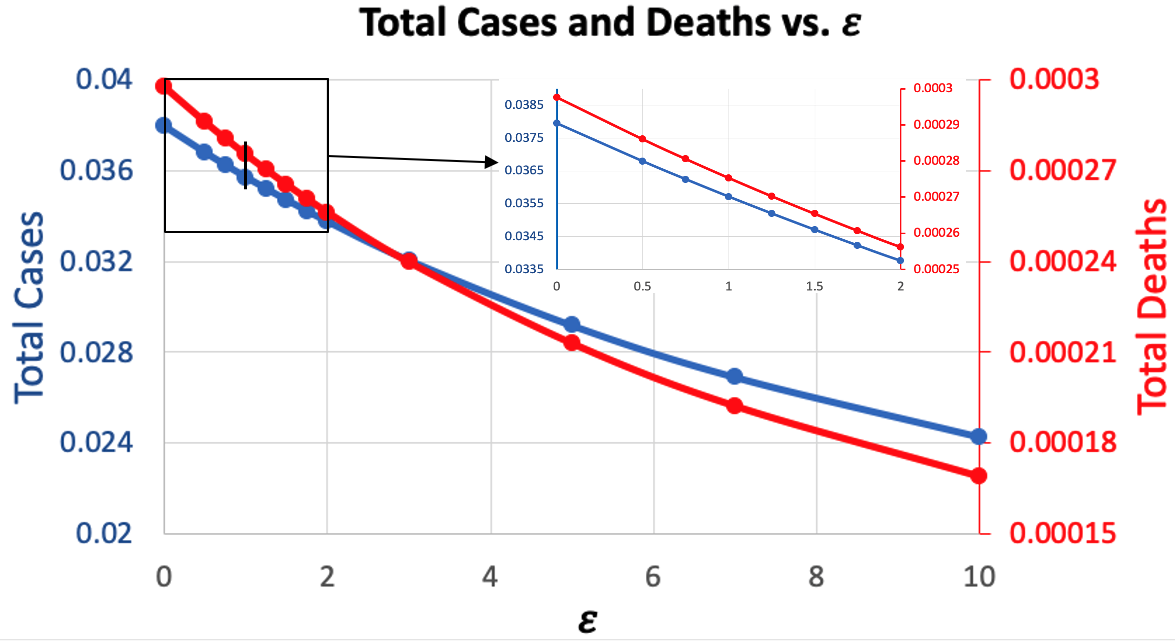}
    \bcaption{Total cases and deaths at different \boldmath$\epsilon$.}{Unlike $\xi$ and $\chi$, the correlation between total cases and deaths and $\epsilon$ appears to be nearly linear. More importantly, testing has a much more insignificant effect on controlling the outbreak. Doubling the testing rate (increasing $\epsilon$ from 1 to 2) only reduces total cases and deaths by about 10\%. Likewise, halving the testing rate only increases total cases and deaths by about 10\%. The minimal effect that testing has may explain why the relationship appears to be linear: the scale is too small for any significant differences to emerge between a linear or exponential relationship. } 
    \label{fig:epsilon}%
\end{figure}

\subsection{Vaccination}

A vaccine with 90\% effectiveness ($\rho = 0.90$) against SARS-CoV-2 has recently been released by industrial pharmaceutical company Pfizer \citep{pfizer}. It is important to investigate the potential effect that this vaccine will have in controlling the outbreak, which will be done by tuning $\theta$, the availability of the vaccine. 
\\
\begin{figure}[H] 
    \subfloat[\textbf{Total cases/deaths by \boldmath$t=380$ at different \boldmath$\theta$.} I notice an exponential relationship between $\theta$ and cases/deaths, with increases in $\theta$ having a more pronounced effect at smaller values of $\theta$. For instance, doubling $\theta$ from 0.005 to 0.01 reduces total cases by approximately twofold, but doubling $\theta$ again to 0.02 only reduces total cases by about 50\%. ]
    {{\includegraphics[width=7.2cm,height=4cm]{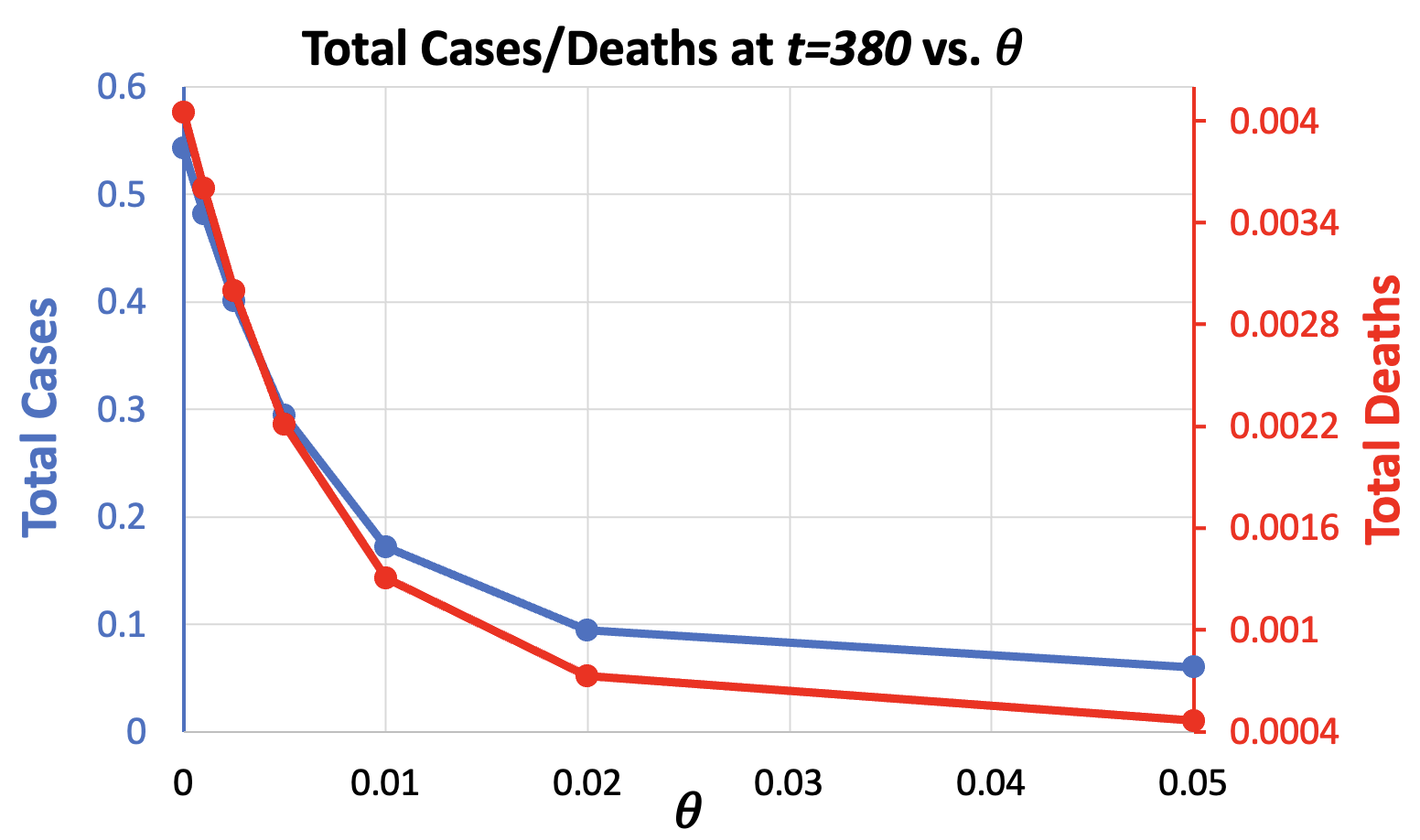} }}%
    \quad
    \subfloat[\textbf{Total cases/deaths by 12/7/20 depending on when the vaccine was released (\boldmath$\theta =0.005$).} The relationship is nearly linear, although changes in the number of days the vaccine is available are slightly more significant at larger values. ]
    {{\includegraphics[width=7.2cm,height=4cm]{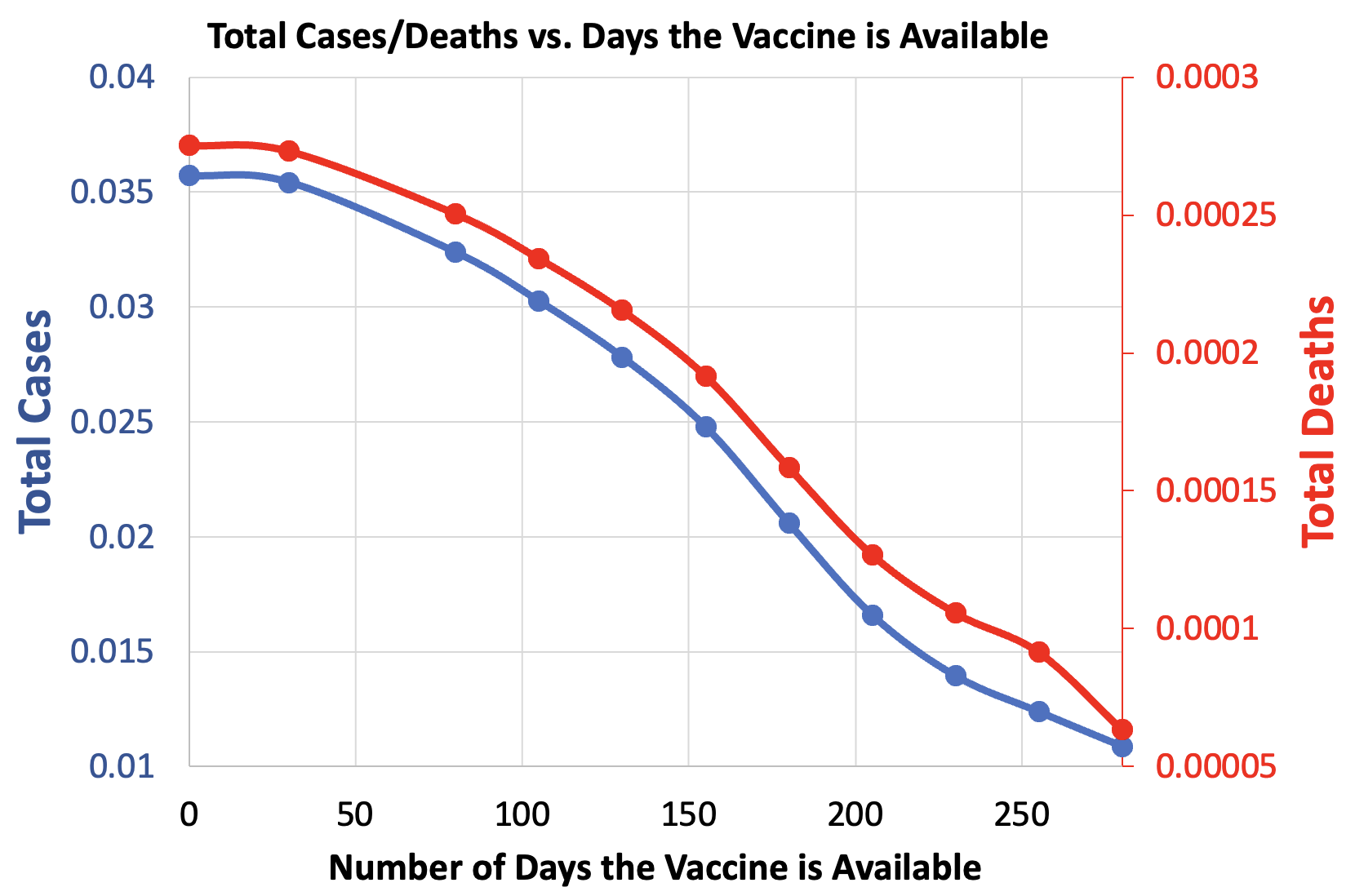} }}%
    \label{fig:vaccine}%
    \caption{\textbf{Total cases and deaths with different vaccine conditions.}}%
\end{figure}

\subsection{Phase Times}

Due to the incorporation of a variable infection rate $\beta_t$ that takes the form of a piecewise function, I can explore the effect that different phase times of $\beta_t$, or different social behavior patterns, would have. For instance, by extending $\beta_{t1}$, a phase of exponential decline in the infection rate, I can examine the potential outbreak that could have occurred if initial social behavior was maintained. 
\vspace{0.5cm}
\begin{figure}[H]
    \centering
    \subfloat[\textbf{Predicted outbreak with an unchanged \boldmath$\beta_t$ and predicted outbreak with \boldmath$\beta_t = \beta_1 $ for all $t$.} The latter scenario reflects one where initial social behavior was preserved throughout the pandemic. In this case, the outbreak would have been over after 75 days. Total cases and deaths would be reduced by over three times.]{{ \includegraphics[width=7.2cm,height=4cm]{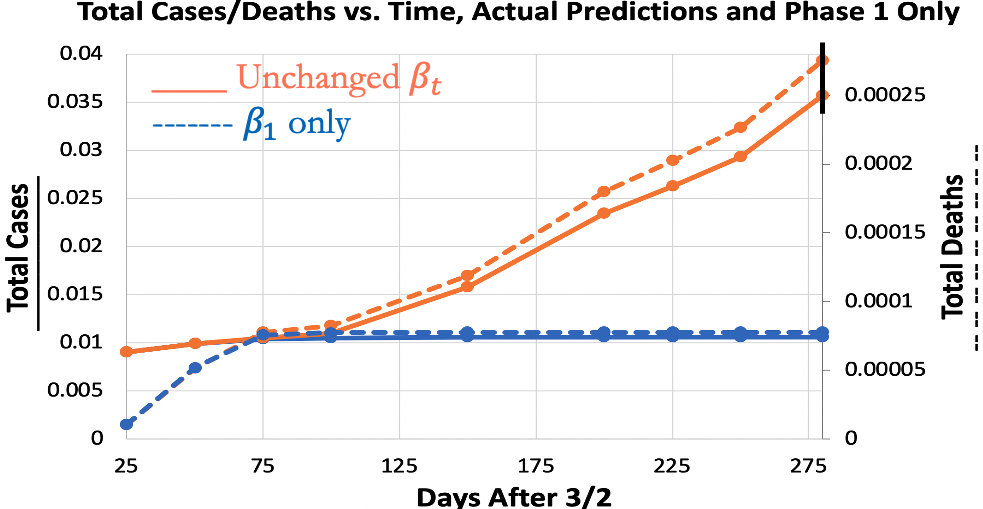} }}%
    \quad
    \subfloat[\textbf{The \boldmath$\beta_t$ functions in Figure \ref{fig:phase times}a.} Because $\beta_1$ is an exponentially decreasing function, by around $t=75$, $\beta_1$ is nearly zero. Although the infection rate would not realistically approach zero, the figure still provides a mechanistic simulation of an outbreak where the infection rate did not increase. ]
    {{\includegraphics[width=7.2cm,height=4cm]{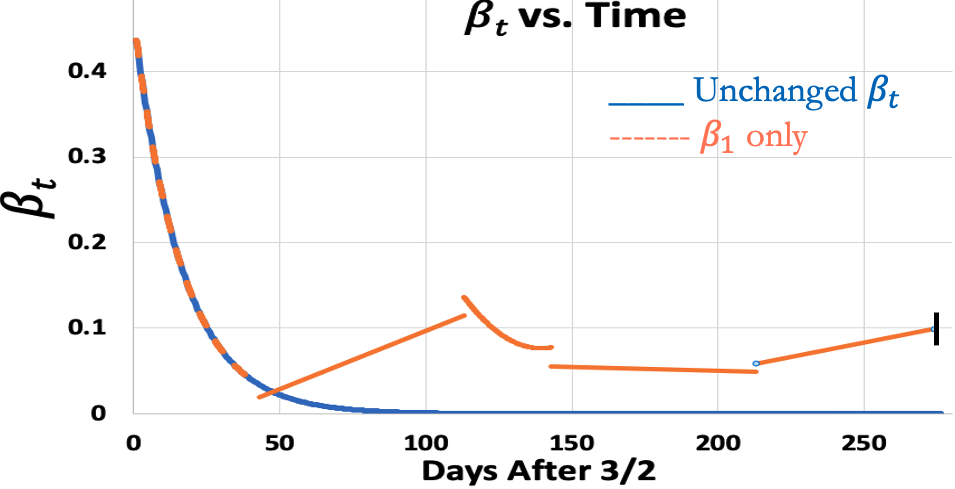} }}%
    \quad
    \subfloat[\textbf{Predicted outbreak with unchanged \boldmath$\beta_t$ vs. \boldmath$\beta_5 = \beta_4 $ for all $t$ vs. \boldmath$\beta_5 = \beta_4$ for $t>280$.} The orange curve reflects the current outlook. The blue curve shows that the brunt of the pandemic could have been avoided if infection rates had not recently risen. However, the green curve demonstrates that the current outbreak is not inevitable; by reverting to the phase four infection rate, the upcoming spike can still be averted.]
    {{\includegraphics[width=7.2cm,height=4cm]{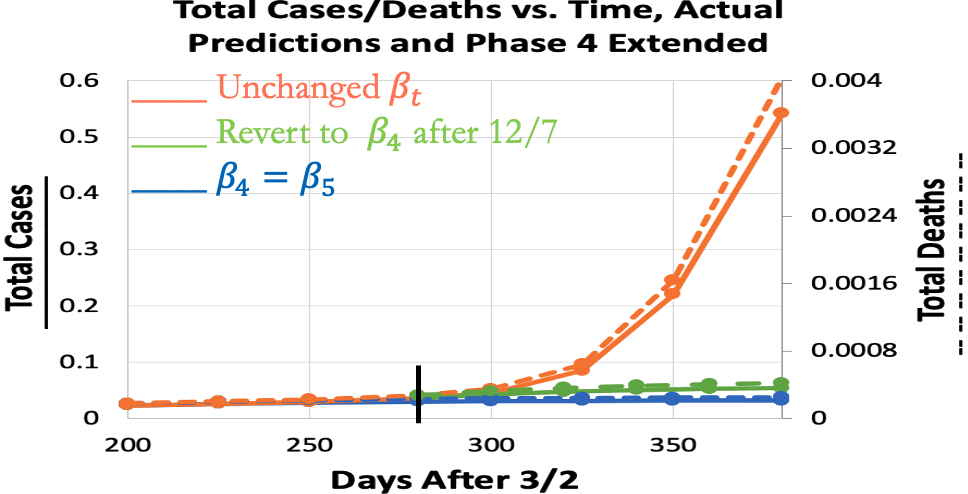} }}%
    \quad
    \subfloat[\textbf{The \boldmath$\beta_t$ functions in Figure \ref{fig:phase times}c.} The blue curve is not an exact calculation, as the infection rate would not continue to decline linearly; likewise, the green curve would not suddenly and discontinuously decline to the value of $\beta_4$ at $t=280$. Still, the analysis provides mechanistic insight into different levels of social behavior.]
    {{\includegraphics[width=7.2cm,height=4cm]{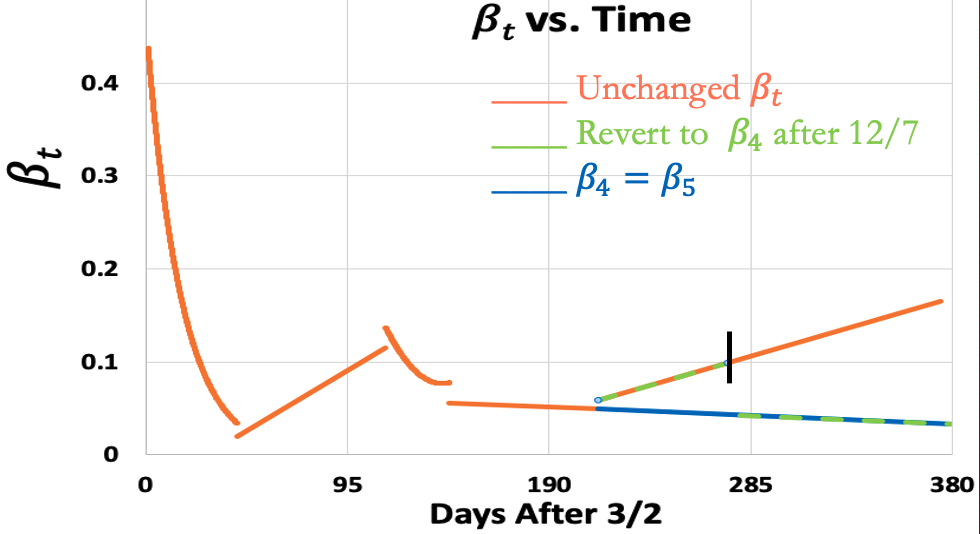} }}%
   \caption{\textbf{Total cases and deaths with different vaccine availabilities}}%
    \label{fig:phase times}%
\end{figure}

\subsection{Summary and Key Findings}

The study of common preventative measures is the most novel and significant aspect of the research. The following tables summarize the primary findings: 

\begin{center}
\begin{table}[H]
\centering
\begin{tabular}{ |c|c|c|c| } 
 \hline
 Parameter & 10\% Increase ($\theta = 0.001$) & 25\% Increase ($\theta = 0.0025$) & 50\% Increase ($\theta = 0.005$) \\ 
 \hline
 $\xi$ & 40.54\% & 64.86\% & 72.97\%\\ 
 \hline
  $\chi$ & 29.73\% & 52.51\% & 60.89\%\\ 
    \hline
  $\theta$ & 11.23\% & 26.34\% & 45.86\% \\ 
 \hline
  $\epsilon$ & 0.56\% & 1.39\% & 2.78\%\\ 
 \hline
\end{tabular}
\caption{ \textbf{A summary of the effect of the preventative measures studied on total cases.} Columns 2-4 show the percent reductions in total cases by $t=280$ given the specified value of $\xi, \chi, \theta$ or $\epsilon$.}
\end{table}
\vspace{-0.5cm}
\end{center}


As discussed in the individual sections and shown in Table 1, the preventative measures studied have varying levels of significance. Reducing the infection rate, which is done mathematically by increasing $\xi$ or $\chi$, is the single  most important method of reducing spread. Translating these findings to the real world, this means that governments and individuals alike must concentrate on social distancing, wearing masks, reducing non-essential travel, quarantining appropriately, etc. Attempting to substitute such measures with widespread testing---which has an almost trivial impact in comparison---would fail to control the spread of the outbreak. Meanwhile, the ongoing distribution of the vaccine must be continued, as public vaccination---like reducing the infection rate---significantly reduces the spread of the virus. 

\section{Other Parameters}

Although other strains of coronaviruses have emerged in the past, COVID-19 has been particularly contagious \citep{sars, mers15}. Some researchers have attributed this to its long incubation period and high rate of asymptomatic cases relative to other diseases \citep{extracontagious}. By tuning the incubation rate $\lambda$ and the asymptomatic rate $\alpha$, I can investigate the significance of these two parameters: 


\begin{figure}[H]
    \centering
    \subfloat[\textbf{Total cases and deaths vs. \boldmath$\lambda$.} There is a negative exponential relationship between $\lambda$ and the spread of the virus. Decreasing the incubation period by just one day would nearly reduce total cases and deaths by twofold. Similarly, increasing the incubation period by just one day would more than double total cases and deaths. These findings corroborate and quantify the significance of the incubation period on the transmission of COVID-19.  ]{{ \includegraphics[width=7.2cm,height=4cm]{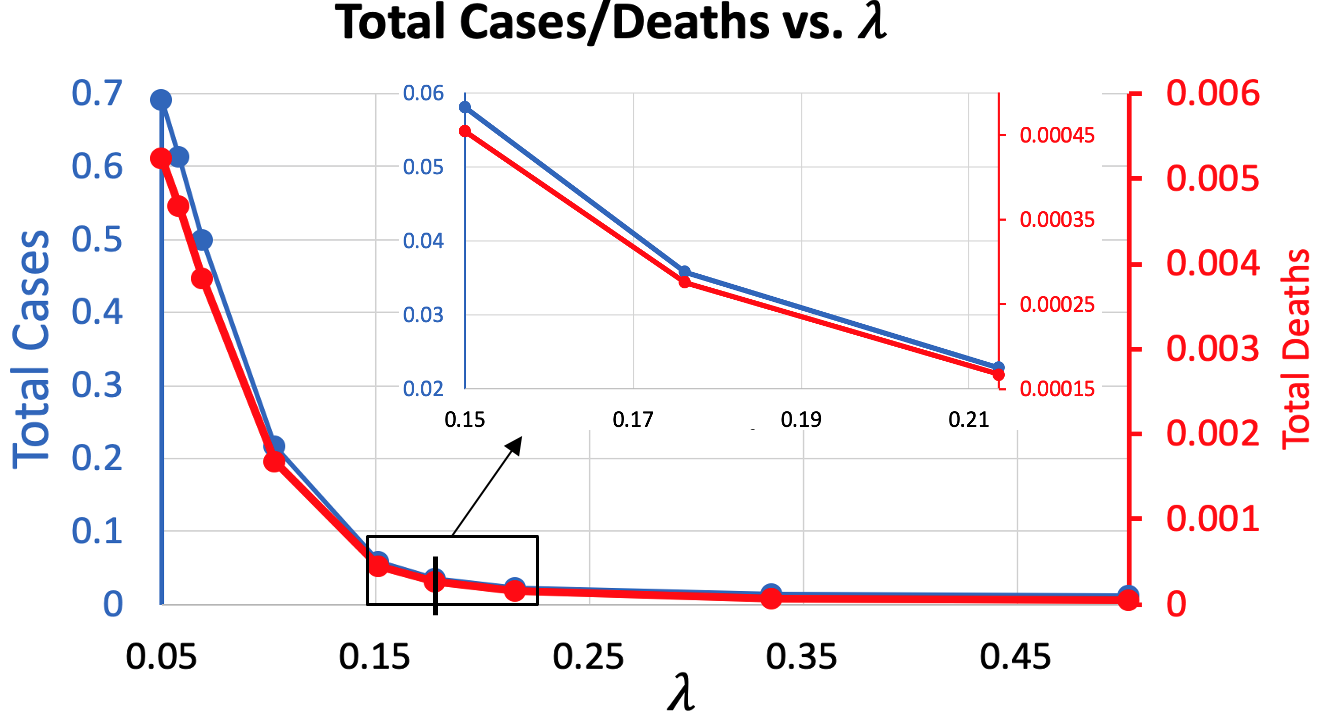} }}%
    \quad
    \subfloat[\textbf{Total cases and deaths vs. \boldmath$\alpha$.} $\alpha$ has a positive exponential relationship with the spread of the virus, with larger values of $\alpha$ leading to a higher number of cases. Increasing the asymptomatic rate from 0.31 to 0.5 would approximately double the total cases while decreasing $\alpha$ to 0.1 would reduce cases by almost twofold.]{{\includegraphics[width=7.2cm,height=4cm]{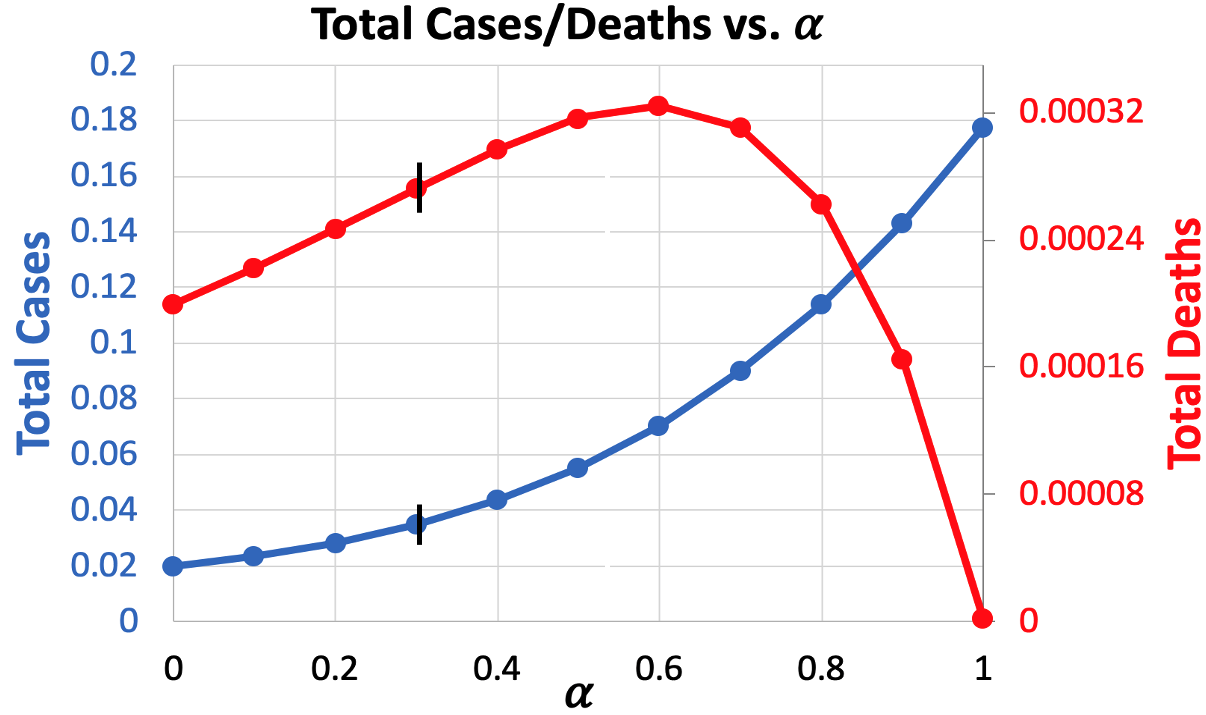} }}%
  \caption{\textbf{Total cases and deaths vs. $\lambda$ and $\alpha$.}}%
\end{figure}

    




 As expected, the total number of cases increases alongside $\alpha$, as a higher rate of asymptomatic infection would lead to more unknown cases, whose infection rates are given by $\beta_{\xi t}$; $\beta_{\xi t}$ is greater than $\beta_{\chi t}$, the infection rate for known, quarantined infections. However, the relationship between total deaths and $\alpha$ is more interesting, as it contains both upwards and downwards-sloping segments. Total deaths increase alongside $\alpha$ until $\alpha \approx 0.6$, making $\alpha \approx 0.6$ the deadliest asymptomatic rate. This correlation can be explained easily: more asymptomatic infections increase the total number of cases, while also creating fewer deadly cases (since only symptomatic infections can cause death). 

\subsection{Death Rate}

Finally, I can investigate the effect of changing the death rate. Notice that there is no parameter incorporated into the model that scales the death rate because I do not wish to tune the entire death rate, but instead only the first two stages, where the death rate was changing rapidly. In stages 3 and 4, the death rate is nearly constant, which I assume to be the equilibrium stage of the death rate where healthcare services are properly equipped to handle the outbreak. Only in the first two stages were these services still adapting to the pandemic, resulting in large fluctuations in the rate of fatal infections. It is these fluctuations that I wish to model. Therefore, in this section, I manually scale the first two stages of the death rate. 

The death rate has a marginal effect on the number of cases since such a small portion of infections are deadly. Thus, it seems only necessary to study the effect of changing the death rate on the actual number of deaths.  

\begin{figure}[H]
    \centering
    \includegraphics[width=14cm,height=5cm]{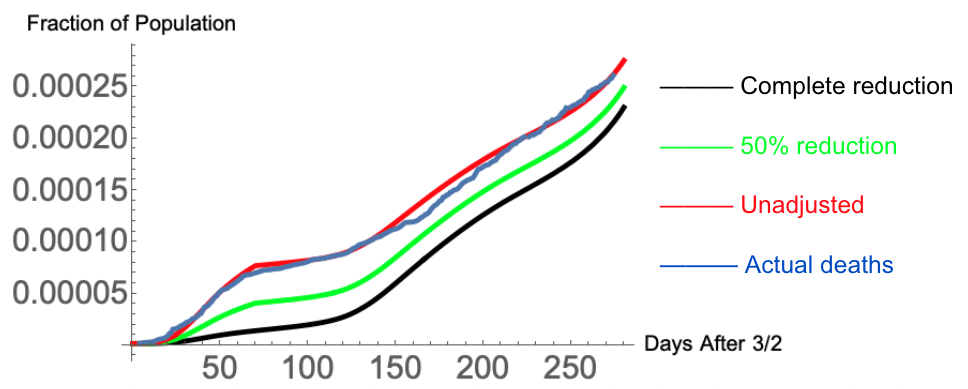}
    \captionof{figure}{\textbf{Predicted vs. actual deaths with different \boldmath$\delta_t$.} Variations in the death rate produce a noticeable but relatively insignificant effect on the number of deaths. The black curve illustrates the course of the outbreak if $\delta_1 = \delta_2 = \delta_{average}$, where $\delta_{average}$ is the average value of the death rate during its third and fourth stages. In other words, it reflects the number of deaths if the death rate had not rapidly spiked at the beginning of the pandemic. The impact is not significant, with deaths decreasing by approximately 20\%. A similar effect is shown by the green curve, which simulates an outbreak where the death rate during the first two phases was reduced by a factor of two; that is, the spike in the death rate was reduced. In this scenario, total deaths were reduced by approximately 10\%. } 
    \label{fig:death rates}%
\end{figure}

Although both situations significantly lower the death rate, such changes do not produce equally significant effects. This is because, during the first two stages, the outbreak was not as widespread, and---although the death rate was higher---there were few cases and therefore few total fatalities. Although the death rate was lower during the third and fourth stages, the brunt of the infections occurred during these phases, thereby producing the majority of the deaths. 

\section{Conclusion}

As the COVID-19 pandemic continues to surge, lives are lost and social and economic repercussions grow worse. Forecasting and mitigating the spread of the pandemic are more important than ever. In this study, I develop a compartmental model specific to the COVID-19 pandemic that accounts for common preventative measures that have been imposed, apply the model to a real-world case study, and use the model to analyze the effect of common preventative measures.

To accomplish this, I derive a novel compartmental epidemiological model that expands on the classic SIR model and accounts for the incubation period, asymptomatic infections, testing, quarantine, vaccination, and death. In addition, I consider the real-world dynamics that lead to variable infection, testing, and death rates. Then, I apply this model to Santa Clara County, California, and, by tuning the model's parameters, I determine the extent to which common preventative measures contain the spread of the virus. Additionally, I study two of the characteristics of SARS-CoV-2 that make COVID-19 particularly contagious. 

\subsection{Key Findings}

I develop a novel SIAPQVRD model that includes eight categories: susceptible ($S$), incubation ($I$), asymptomatic ($A$), asymptomatic and quarantined due to a positive test ($P$), symptomatic and quarantined($Q$), recovered ($R$), vaccinated ($V$), and dead ($D$). In addition, I include variable infection, testing, and death rates.

To apply the model to Santa Clara County, I take publicly-released COVID-19 data and derive functions for the infection, testing, and death rates. I find that the infection rate can be classified into five distinct phases, reflect changes social interactions over time. The testing rate rises consistently over time and can also be divided into five stages that each represent different rates of growth. The death rate can be divided into four stages, which reflects the adaptation of healthcare services. The remaining parameters were determined using external research on COVID-19, except $\chi$, which was found experimentally. 

I first compare the model's predictions with actual data by inputting initial conditions and find that it is reasonably accurate, albeit with minor discrepancies. Then, to investigate the effect of preventative measures, I tune parameters in the model individually while holding the others constant, making the tuned parameter the independent variable and the total number of cases and deaths by the time of writing the dependent variable. I find that controlling the infection rate through parameters $\xi$ and $\chi$ has the most significant impact on containing the virus. A vaccination also has a significant effect on reducing viral transmission, with a more available vaccine and a greater duration of availability producing a substantially stronger effect. Testing, however, barely has a noticeable impact, even at very high levels. By experimenting with different phase times for the infection rate, I also find that reducing social behavior to prevent the infection rate from rising is critical to minimize transmission. Indeed, reverting to initial social interaction levels could end the pandemic shortly.

I also investigate the effect of tuning the incubation period and rate of asymptomatic infections, and I find that both have a significant effect. Indeed, the long incubation period and high rate of asymptomatic cases relative to other diseases appear to be driving causes of COVID-19's rapid spread. Decreases in the incubation period exponentially decrease total cases and deaths, while increases in the asymptomatic rate also exponentially increase total cases. The asymptomatic rate increases the number of deaths for $\alpha \lesssim 0.6$ and decreases the number of deaths for $0.6 \lesssim \alpha \leq 1$

\subsection{Implications and Applications of the Findings}
\subsubsection{General SIAPQVRD Model}

The model is a novel development in compartmental epidemiological modeling. Although it was designed specifically for COVID-19, many infectious diseases also involve a significant incubation period and cause asymptomatic infections, such as the H1N1 influenza virus or MERS-CoV \citep{MERSSpread, H1N1}. In addition, any modern pandemic will also involve the implementation of the preventative measures studied---social distancing, quarantine, widespread testing, and vaccination. Thus, the model is not exclusive to COVID-19 and lays a paramount foundation for the mathematical modeling of forecasting future pandemics. 

\subsubsection{Case Study of Santa Clara County, CA} 



First, the observation that $\chi = 1.75$ shows that quarantined individuals still have a reasonably high infection rate, creating only approximately 1.75 fewer secondary infections per day than unknown, asymptomatic infections. This proves that individuals who know they are infected and contagious still maintain relatively high levels of social interactions. In an ideal situation with a perfect quarantine, $\chi \rightarrow \infty$. I find that increasing $\chi$ to a sufficiently high number could reduce total cases and deaths by over three times. The implication is twofold: first, that individuals who are symptomatic or have been tested positive for COVID-19 must quarantine themselves more effectively; second, that governments must enforce stricter regulations that mandate the quarantine of anyone with a known infection. 


Second, I discover that the death rate rises and then falls rapidly at the beginning of the outbreak. Afterward, it remains nearly constant. This indicates that hospitals and other public services are initially unprepared to handle surges in cases. As such facilities adapt, despite increases in the number of cases, they treat patients more effectively, thereby reducing the death rate. I find that if hospitals were better-equipped initially and the death rate during the spike was reduced twofold, total deaths could be reduced by $~10\%$. In an ideal situation where such services were fully prepared from the beginning (i.e. the death rate was at the equilibrium value from the start), total deaths could be reduced by $~20\%$. Although such reductions would not be particularly significant in Santa Clara County, where there have been few COVID-19 deaths altogether, they could have much more meaningful effects in areas hit harder by the pandemic.

\subsubsection{Study of Preventative Measures}

My study of key preventative measures has the most real-world implications. First, I notice that tuning $\xi$ and $\chi$, which scale the infection rate, has a dramatic effect on the outbreak, with minor increases in either parameter resulting in very significant declines in the number of cases and deaths. Therefore, governments' first priorities should be to minimize the infection rate. Although I reference $\xi$ as the social distancing parameter throughout the study, it more generally refers to the regular social patterns of all individuals; therefore, any preventive measure that is adopted by all individuals such as mask-wearing or minimizing non-essential travel would increase $\xi$. To increase $\chi$, a stronger quarantine for those who know they are infected should be enforced. 

I also find that if social behavior followed the same pattern as it did during the first stage of the infection rate, the pandemic in Santa Clara County would have ended in just 75 days. The upcoming spike can also be avoided by reverting to previous levels of social patterns, where the infection rate declined linearly, albeit slowly. Therefore, I find that the number one method to reduce viral transmission is to ensure that the infection rate does not rise; even minor decreases in the infection will, over time, force an end to the outbreak. In the scope of the model, this makes sense: if the infection rate increases linearly, then the rate at which people spread the disease increases linearly, causing the number of infections to rise exponentially. 

Unlike the infection rate, testing has a much smaller effect on controlling the outbreak, with a twofold increase in the testing rate reducing total cases and deaths by $10\%$. The implication is first that testing will never reach adequately significant levels to have a meaningful reduction in transmission; and, second, that even with high testing rates, testing is insignificant because those who receive a positive test do not quarantine themselves adequately. Thus, governments should consider channeling resources away from testing towards increasing other preventative measures that could have a more significant impact on reducing the outbreak. Meanwhile, governments should impose regulations or spread information that results in stronger quarantines; if quarantines are adequately effective, then testing can have a stronger preventative impact, as those with an asymptomatic infection who receive a test will spread the virus at greatly reduced rate. 

The distribution of the vaccine will have a significant impact on reducing the spread of the virus. Increasing the availability of the vaccine is particularly important, especially at lower availabilities. Therefore, alongside reducing the infection rate, governments should prioritize distributing the vaccine as fast as possible.

Interestingly, the death rate has a small effect on reducing the death count. Even if public services such as hospitals were properly equipped from the beginning, total deaths would have been reduced by only 20\%. This is because most infections occurred once the death rate had stabilized. Therefore, I can conclude that although ill-prepared initially, such services reacted in adequate time to ensure that the death rate was reduced for the majority of the outbreak. Therefore, although better-equipped hospitals can save a moderate number of lives, in future pandemics, governments can direct more attention to reducing the number of infections, which is the true driver of fatalities. 

I also find that longer incubation periods significantly increase contagion. Although the incubation period of the virus cannot be changed, this result emphasizes that individuals must quarantine themselves adequately if they have been in contact with an infected individual, as they may be carriers of the virus without exhibiting symptoms. Likewise, this demonstrates that someone who develops symptoms for COVID-19 could have already spread the virus to many people in the days prior. Similarly, the high asymptomatic rate of COVID-19 is a driving cause of the disease's high infectivity, which reinforces that individuals must take adequate precautions under the possibility of being infected. However, the deadliest asymptomatic rate occurs at $\alpha \approx 0.6$, where the rate is high enough to drive rapid spread but not so much so that few cases are deadly.  

\subsection{Limitations and Considerations of the Model}

As with all epidemiological models, there are limitations of the analysis. First, I will address the model's primary assumptions and simplifications.

\begin{enumerate}
    \item The law of mass action is the most fundamental assumption made in all compartmental modeling, but it can still lead to errors. In reality, not all individuals have equal probabilities of interaction. For example, those who are vaccinated first may be a part of at-risk groups who already have infrequent interactions with other individuals. There are countless other scenarios where this assumption fails. However, because it is widespread and unavoidable in disease modeling, it should be considered as a natural and expected source of error. 
    \item A fixed population is unlikely to have a significant effect on the accuracy of the model. In the context of a pandemic, even in a worst-case scenario (e.g. five years), the population would not change dramatically, and any models would still be almost entirely unchanged. 
    \item The assumption of widespread social distancing and quarantine can also skew the results. Of course, not all individuals will obey social distancing restrictions or quarantine adequately, a trend that has been independently studied \citep{distrust}. However, it would be extremely difficult to both identify the exact effect of this assumption and account for it. Like the law of mass action, it is a source of natural error in epidemiological modeling. One method to mitigate its impact could be a subcompartmentalization of categories, for instance by dividing each infected population into smaller groups that have scaled infection rates.
    \item The simplification to only track the vaccination of susceptible individuals is unlikely to significantly alter the accuracy of the model. An infected or recovered individual is very unlikely to receive a vaccination, given that they are ultimately a very small proportion of the population; the vaccine would also have little to no effect, since the individual would either soon or already be immune to COVID-19. 
    \item The assumption that vaccination produces immediate effects can also be a source of error. In reality, the Pfizer vaccine requires two separate shots, with the first dose only providing 52\% efficacy \citep{pfizer2}. The second vaccine should be given after approximately three weeks, and this booster shot leads to the 90\% efficacy threshold that was found and used in my analysis \citep{twodose}. In addition, it takes roughly one month after the initial shot for the effects of the vaccine to materialize \citep{onemonth}. Therefore, the impact of vaccination in this study may be overstated: in reality, the vaccine both requires multiple doses and does not produce an immediate effect. To account for this, I could divide the vaccinated compartment into two subclassifications depending on how many doses of the vaccine an individual has received. In addition, I could implement a delay between an individual's vaccination and their immunity. 
    \item The simplification to only track the testing of asymptomatic individuals may also generate error. I argue that it is only relevant to consider testing on asymptomatically infected individuals because a negative test would not change an individual's behavior (and therefore, the infection rate). I can also assume that all symptomatic individuals are tested, but this test is unlikely to change their social behavior as they would already be quarantining (and therefore, it is not necessary to create a separate compartment for these individuals). However, in reality, a negative test may increase one's social behavior because they are not afraid of transmitting the disease to others. Therefore, the model may underestimate the severity of the pandemic because testing could either increase the infection rate in the event of a negative test. A potential improvement to the model could involve tracking the testing of all individuals and incorporating another compartment for those who receive a negative test.
    \item The assumption of perfect testing can lead to errors, as COVID-19 testing is occasionally inaccurate. Studies have found that the rate of false negatives can be as high as 29\% \citep{falseneg}. The false-positive rate is, on average, between 0.8\% and 4.0\%, but can be much higher, reaching over 50\% in scenarios where the people being tested were previously in close contact with infected individuals \citep{falsepos1, falsepos2}. Since the false-negative rate is higher than the false-positive rate, the impact of testing may be overstated. To account for this, I could include parameters that dictate the success rate of the test, as I do for vaccination, to randomly create false positives or false negatives. However, as I find that testing ultimately has a minimal impact on the outbreak as a whole, this assumption likely does not dramatically affect the accuracy of the predictions. 
\end{enumerate}

The process of calculating parameter values also produces inaccuracies. As discussed in detail in their respective sections, I make notable assumptions in calculating the infection rate and the death rate. In determining the infection rate, I make a SIR-model approximation and assume that the infection rate can be given by the number of new cases divided by the product of the susceptible population and the number of reported cases. To account for this approximation, I choose an appropriate value of $\chi$ that makes the model as accurate as possible. However, this is not necessarily the most accurate way to approximate the infection rate, since the infection rate for quarantined individuals may be more complex than just a scalar multiple of the infection rate for non-quarantined individuals. It would be difficult to determine the exact adjustment, though, since it is impossible to know the precise number of asymptomatic individuals and their exact infection rates. 

The death rate is also an estimation because a fraction of the reported cases are asymptomatic and cannot lead to death, which I ignore. To account for this, one could use the model to estimate how many of the reported cases at any given time are asymptomatic, subtract that population from the number of reported cases, then use the adjusted difference to calculate the death rate. However, the assumption is mostly insignificant, as verified by the fact that the death rate prediction was extremely accurate and that there are few known asymptomatic cases at all times. 


\subsection{Future Explorations}

As a whole, one major limitation of the model is that I need data to produce data; that is, I was only able to perform the analysis of common preventative measures because I already had data that allowed us to calculate functions for the infection rate, testing rate, and death rate. Therefore, a future exploration in this field would be to predict such rates for the future and forecast the outbreak by inputting such rates into the model. Indeed, research into statistical physics is already being conducted amid the pandemic to predict future infection and recovery rates \citep{predictfuturerates}. Biological research into the possibility of new viral strains or mutations---both of which have been observed empirically in many contagious diseases, including coronaviruses---would also facilitate the accuracy of the model by predicting when key parameter values may change in the future \citep{Strain20,Sa16}. As the model very accurately predicts the spread of the disease when given appropriate parameter values, a combination of this model and accurate projections of future parameter values could prove very useful in forecasting future outbreaks. 

The model itself could also be advanced. Future research could include intercompartmental classifications, such as by age, gender, or health status to more accurately determine which groups are more at risk of becoming infected or dying from the virus. To better model real-world dynamics, other parameters, such as the recovery or vaccination rate, could also become variable. Non-constant parameters could also be elevated with an infection-age structured epidemic model such as the one proposed by \citet{hyman} that takes into account how long an individual has been infected to calculate the values for their parameters (e.g. an individual who has been infected for longer has a lower transmission rate due to decreased interactions). The model could further be improved by adding a stochastic perturbation that would better mirror the random probabilities of real-world dynamics, such as what \citet{SIVR} investigate in their derivation of the SIVR model. 
\pagebreak
\section*{Acknowledgements}

I would like to extend deep gratitude to Ying-Jen Yang (University of Washington) for his support in verifying the mathematical accuracy of the model and its equations, as well as for his contributions in improving the visual quality of the figures. I also thank Anu Aiyer (The Harker School) for providing feedback and suggestions on the writing of the paper, primarily in sections 4 and 7. 

\pagebreak
\printbibliography
\end{document}